\documentclass[%
 reprint,
 amsmath,amssymb,
 aps,
]{revtex4-2}
\usepackage{graphicx}
\usepackage{dcolumn}
\usepackage{bm}
\usepackage{hyperref}
\usepackage{cleveref}
\usepackage{changepage}
\usepackage[labelformat=empty, position=top]{subcaption}
\usepackage[export]{adjustbox}
\usepackage{caption}

\usepackage[T1]{fontenc}
\usepackage[utf8]{inputenc}
\usepackage{floatrow}
\crefname{equation}{Eq.}{Eqs.}
\Crefname{equation}{Eq.}{Eqs.}

\usepackage{comment}

\DeclareUnicodeCharacter{2212}{\textendash}

\begin{document}

\preprint{APS/123-QED}

\title{Synaptic motility and functional stability in the whisker cortex
}

\author{Nimrod Sherf$^{1,2,4}$}
\email{sherfnim@post.bgu.ac.il}
\author{Maoz Shamir$^{1,2,3}$}%
\email{shmaoz@bgu.ac.il }

\affiliation{%
$^1$Department of Physics, $^2$Zlotowski Center for Neuroscience and $^3$Department of Physiology and Cell Biology, Ben-Gurion University of the Negev, Beer-Sheva, Israel, $^4$Department of Mathematics, University of Houston, Houston, Texas, USA. \\
}%




\date{\today}

\begin{abstract}
The high motility of synaptic weights raises the question of how the brain can retain its functionality in the face of constant synaptic remodeling. Here we used the whisker system of rats and mice to study the interplay between synaptic plasticity (motility) and the transmission of sensory signals downstream.

Rats and mice probe their surroundings by rhythmically moving their whiskers back and forth. The azimuthal position of a whisker can be estimated from the activity of whisking neurons that respond selectively to a preferred phase along the whisking cycle. These preferred phases are widely distributed on the ring.  
However, simple models for the transmission of the whisking signal downstream predict a distribution of preferred phases that is an order of magnitude narrower than empirically observed. Here, we suggest that synaptic plasticity in the form of spike-timing-dependent plasticity (STDP) may provide a solution to this conundrum.  This hypothesis is addressed in the framework of a modeling study that investigated the STDP dynamics in a population of synapses that propagates the whisking signal downstream.

The findings showed that for a wide range of parameters, STDP dynamics do not relax to a fixed point. As a result, the preferred phases of downstream neurons drift in time at a non-uniform velocity which in turn, induces a non-uniform distribution of the preferred phases of the downstream population. This demonstrates how functionality, in terms of the distribution of preferred phases, can be retained not simply despite, but because of the constant synaptic motility. Our analysis leads to several key empirical predictions to test this hypothesis.

\end{abstract}

\keywords{Spike-Timing-Dependent Plasticity}
\maketitle


\section{\label{intro}Introduction}
Mice and rats scan their proximal environment by moving their whiskers back and forth in a rhythmic manner  \cite{sachdev2002divergent,wallach2020predictive,knutsen2006haptic,mehta2007active,landers2006development,hill2011primary}. 
Whisking provides important tactile signals that allow rats and mice to detect, locate, and identify nearby objects  \cite{isett2020cortical,kleinfeld2011neuronal,ahissar2008object,diamond2008and}. The tactile signal from the whiskers is transduced to a series of short electrical pulses (spike trains) that propagate downstream the whisker system: from the trigeminal ganglion to the brain-stem, to the thalamus to layer 4 of the cortex downstream to layer 2/3 \cite{kleinfeld2011neuronal,diamond2008and,yu2006parallel,house2011parallel,yu2016layer,petersen2019sensorimotor,voelcker2022transformation}, \hyperlink{fig 1}{Fig.\ 1(a)}.

Throughout the whisker system, neurons that encode the \emph{phase} along the whisking cycle have been found \cite{yu2006parallel,diamond2008and,severson2017active,isett2020cortical}. These neurons, termed whisking neurons, fire in a selective manner to the phase along the whisking cycle, \hyperlink{fig 1}{Figs.\ 1(b)-1(c)}. Typically, whisking neurons are characterized by a unimodal tuning curve that peaks at the preferred phase of the neuron  \cite{ego2012coding,moore2015vibrissa,yu2016layer,grion2016coherence,severson2017active,yu2019recruitment,gutnisky2017mechanisms,isett2020cortical}, \hyperlink{fig 1}{Figs.\ 1(d)-1(e)}.

The preferred phases of whisking neurons are widely distributed on the ring in a non-uniform manner, \hyperlink{fig 1}{Fig.\ 1(g)}. The distribution of preferred phases in specific brain regions can be approximated by a von Mises (circular normal) distribution, 
\begin{equation}
    \label{eq:VM4}
    \Pr (\phi) = \frac{e^{\kappa \cos(\phi-\psi)}}{2 \pi I_{0}(\kappa)}
\end{equation}
where $I_0(\kappa)$ is the modified Bessel function of order 0. The parameters $\psi$ and $\kappa$ quantify the mean and width (or sharpness) of the distribution. Typically, the value of $\kappa$ is around 1, whereas the mean, $\psi$, depends on the brain region \cite{kleinfeld2011neuronal,yu2016layer}. 
 
The distribution of preferred phases poses a challenge to our understanding of the propagation of sensory signals in the central nervous system. To delve into the source of this conundrum, we consider a simple model for the transmission of whisking signals from layer 4 downstream to layer 2/3 (L2/3) in the whisker cortex. In this simplistic model, we ignore the effect of recurrent connectivity within L2/3, as well as the possible contribution of direct thalamic input \hyperlink{fig 1}{Fig.\ 1(g)}. Excitatory neurons in layer 4 of the whisker cortex almost do not respond to whisking, whereas about $30 \%$ of the inhibitory neurons in layer 4 (L4I) are whisking neurons \cite{peron2015cellular,peron2020recurrent}.
Thus, the main source of the whisking signal to layer 2/3 originates from L4I neurons.

We model the activity of the downstream L2/3 neuron by a delayed linear response to its inputs (see \cite{,luz2012balancing}). This higher level of abstraction (compared to the frequently used linear non-linear Poisson model \cite{softky1993highly,shadlen1998variable,masquelier2009oscillations,luz2016oscillations,song2000competitive,kempter2001intrinsic,kempter1999hebbian}) is applied to facilitate analysis, and is expressed as
\begin{equation}
\label{eq:postfiring3}
    \rho_{L2/3}(t)=I_{ex}-\frac{1}{N}\sum_{k=1}^{N}w_k \rho_{k}(t-d),
\end{equation} 
where $I_{ex}$ denotes an excitatory drive that is independent of the whisking phase, $w_k$ is the synaptic weight of the $k$th L4I neuron and $d>0$ is the delay. 
Parameter $\rho_k$ denotes the activity of neuron $k$ of the population of L4I neurons that serve as feed-forward input to the downstream L2/3 neuron.
We further assume that the preferred phases of the upstream population, $ \{ \phi_k \}_{k=1}^N$, are i.i.d.\ according to \cref{eq:VM4} with mean $\psi_{ \mathrm{L4I}}$ and width $\kappa_{ \mathrm{L4I}}$.
Thus, the preferred phase of the downstream neuron is determined by a weighted average of the preferred phases of its inputs, weighted by their synaptic strengths.


  \begin{figure*} \hypertarget{fig 1}{}
\begin{adjustwidth}{-2em}{0em}
\begin{subfigure}[t]{0.007\textwidth}\vspace{-9.50cm}\hspace{-0.5cm}
		\textbf{(a)} 
	\end{subfigure}
  \begin{subfigure}[b]{0.40\textwidth}
    \includegraphics[width=\textwidth]{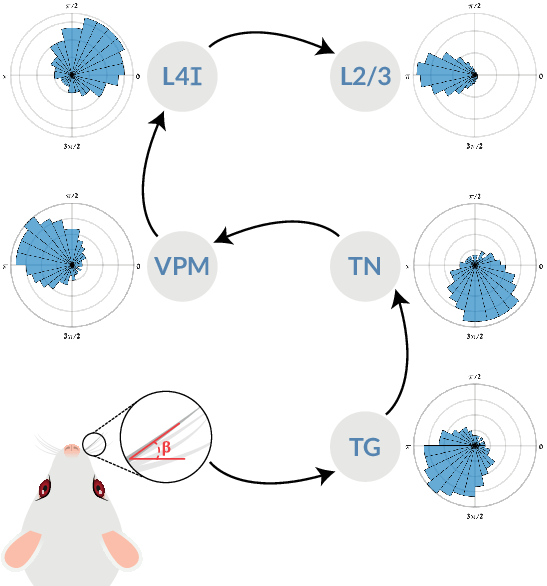}
    \caption{}
    \label{fig:}
  \end{subfigure}
  \hspace{0.5cm}
  \begin{subfigure}[b]{0.70\textwidth}
    \includegraphics[width=\textwidth]{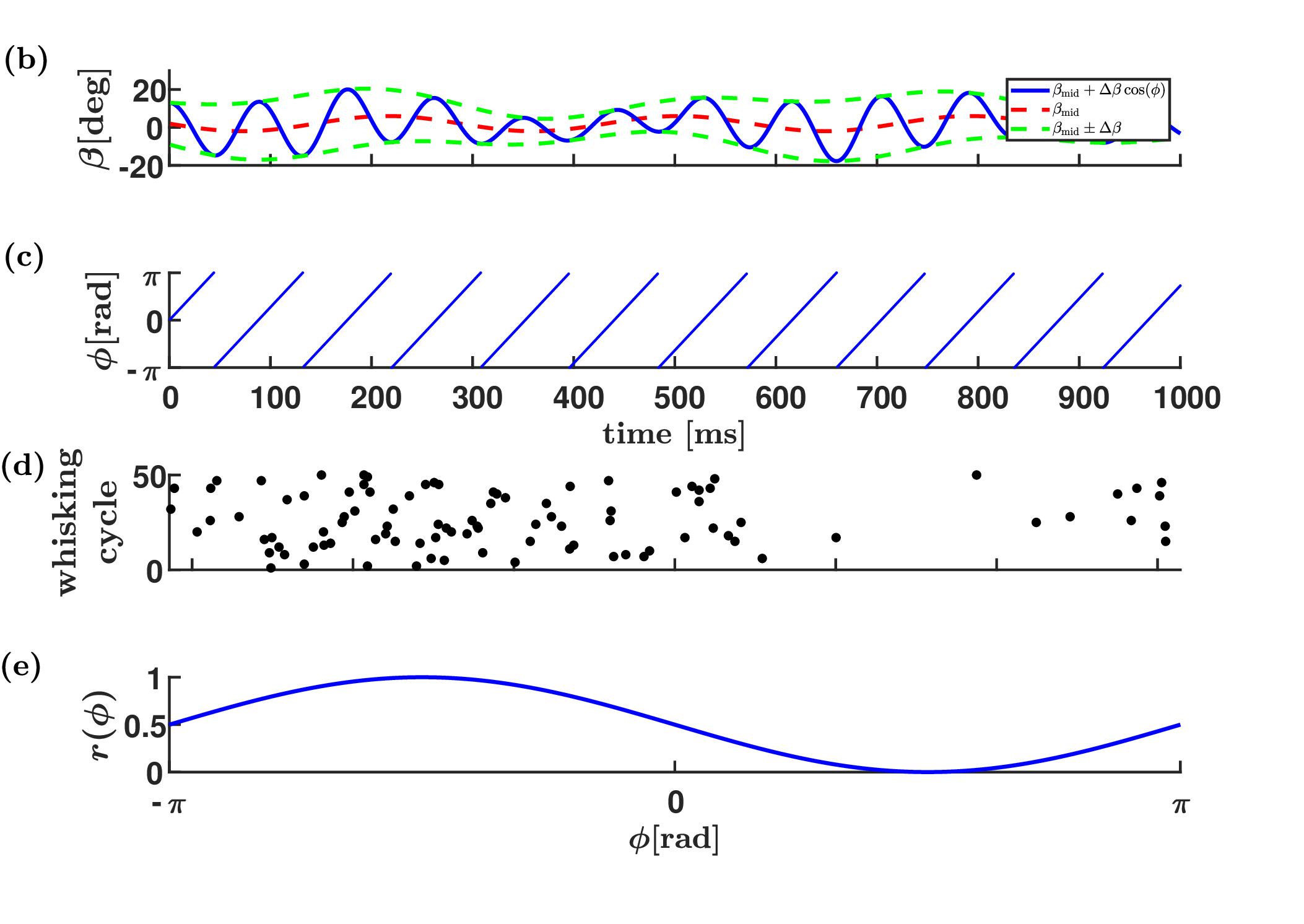}
    \caption{}
    \label{fig:}
  \end{subfigure}
\end{adjustwidth}

\begin{adjustwidth}{-2em}{0em}
\vspace{-6.50cm}
\begin{subfigure}[t]{0.007\textwidth}\vspace{-7.5cm}\hspace{-0.5cm}
		\textbf{(f)} 
	\end{subfigure}
  \begin{subfigure}[b]{0.45\textwidth}\vspace{-4.50cm}
    \includegraphics[width=\textwidth]{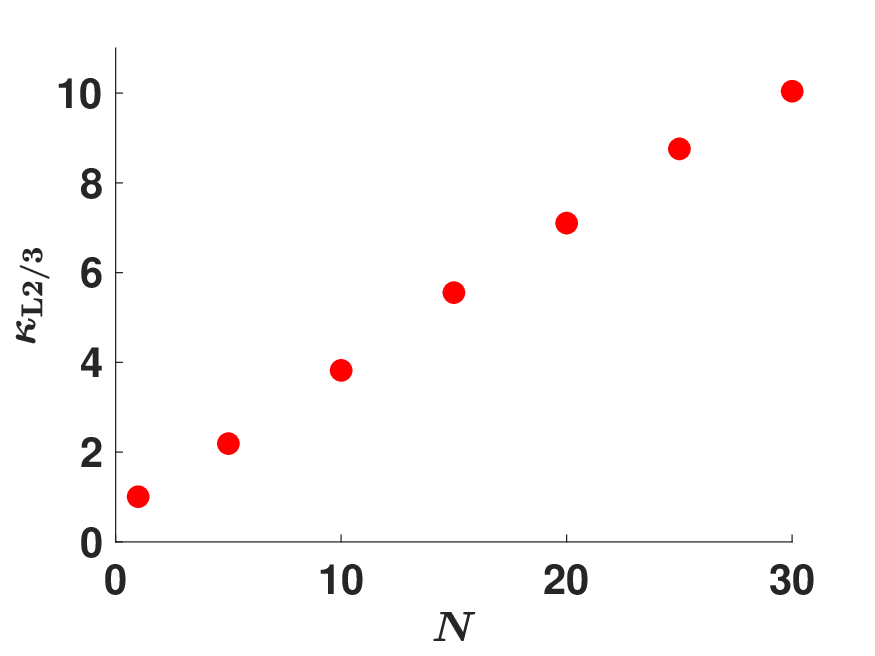}
    \caption{}
    \label{fig:}
  \end{subfigure}
  \hspace{2.5cm}
  \begin{subfigure}[t]{0.007\textwidth}\vspace{-7.50cm}\hspace{-0.5cm}
		\textbf{(g)} 
	\end{subfigure}
  \begin{subfigure}[b]{0.35\textwidth}
  \vspace{5.50cm}
    \includegraphics[width=\textwidth]{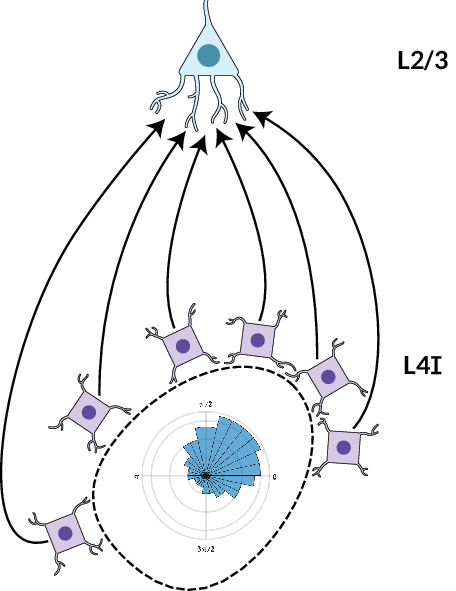}
    \caption{}
    \label{fig:}
  \end{subfigure}
\end{adjustwidth}  \vspace{-0.8cm} 
\caption{ {\bf Transmission of the whisking signal}. (a) In the lemniscal pathway \cite{diamond2008and,yu2006parallel,petersen2019sensorimotor} the whisking signal is relayed downstream via the trigeminal ganglion (TG) to the trigeminal nuclei (TN) of the brainstem down to the ventral posterior medial (VPM) nucleus of the thalamus. From the thalamus, the whisking signal in the lemniscal pathway is transmitted downstream to layer 4 of the cortex (L4I - mainly inhibitory neurons), and then down to layer 2/3 (L2/3) \cite{house2011parallel,yu2016layer,petersen2019sensorimotor,voelcker2022transformation}. The polar histograms represent the distributions of preferred phases along the pathway, based on \cite{isett2020cortical,moore2015vibrissa,yu2016layer,severson2017active} - which are presented purely for purposes of illustration. Note, that the whisking signal is also transmitted via the paralemniscal pathway \cite{diamond2008and,moore2015vibrissa,bureau2006interdigitated}.
(b)-(e) encoding of the whisking signal by a whisking neuron - illustration. (b) The angular position of a whisker, $\beta$, is shown as a function of time. The angle is often modeled as $\beta(t) = \beta_{\text{mid}}(t) + \Delta \beta(t) \cos(\phi(t))$, where $\beta_{\text{mid}}(t)$ and  $\Delta \beta(t)$ are the midpoint and the
whisking amplitude, respectively, that change slowly in time. (c) The whisking phase $\phi$ as a function of time is $\phi(t) = (\nu t)_{\text{mod} 2\pi}$, where $\nu$ is the angular frequency of the whisking. (d) Raster plot. The phases in which a single model whisking neuron spiked are marked (black dots) for different whisking cycles (on the y-axis)). The preferred phase of the model neuron at $\phi=\pi/2$. (e) Tuning curve. The normalized mean firing rate of the model neuron in (d) is shown as a function of the phase along the whisking cycle. (f) The expected width in the downstream layer, $\kappa_{ \mathrm{L2/3}}$, in the random pooling model, is depicted as a function of the number of pooled L4I neurons, $N$, for random pooling. The width of the distribution, $\kappa_{ \mathrm{L2/3}}$, was estimated from 10000 repetitions of drawing $N$ preferred phases of the upstream population with $\kappa_{ \mathrm{L4I}}=1$. (g) Feed-forward architecture of the L4I-L2/3 section of the pathway.}
				\label{fig:intro}
\end{figure*}


However, synaptic weights are highly volatile \cite{fox2002anatomical,erzurumlu2012development,ziv2018synaptic,hazan2020activity}. Fluctuations of about $50 \% $ in the synaptic weights were shown to occur over a period of several days \cite{ziv2018synaptic,hazan2020activity}. Additionally, \emph{activity-independent} fluctuations were reported to be on the same order of magnitude \cite{hazan2020activity}. If synaptic plasticity is purely activity-independent, the synaptic weights,  $ \{ w_k \}_{k=1}^N $, will be random and independent of the preferred phases of the upstream L4I neurons. Thus, the preferred phase of a downstream L2/3 neuron is determined by random pulling of the preferred phases of the $N$ L4I neurons that serve as its input. 
As a result, the width of the distribution of the preferred phases of L2/3 whisking neurons will decay to zero as the number of pooled L4I neurons, $N$, grows, \hyperlink{fig 1}{Fig.\ 1(f)}. Even pooling responses of as few as $N=20$ L4I neurons results in $\kappa_{\mathrm{L2/3}} \approx 10$, which is an order of magnitude larger than typically observed \cite{sherfshamir2021}. Raising the question: what mechanism can generate this distribution?

Recently, we studied the transmission of whisking signals from excitatory neurons in the thalamus downstream to layer L4I neurons. We showed that activity-dependent plasticity in the form of spike-timing-dependent plasticity (STDP, as detailed in \cref{STDPmodel}) can provide a mechanism that generates a non-trivial distribution of preferred phases governed by the STDP rule \cite{sherfshamir2021}. However, this work only considered the plasticity of excitatory synapses and capitalized on the strong positive feedback inherent to excitatory STDP \cite{shamir2019theories}. In contrast, inhibitory STDP is typically characterized by a negative feedback mechanism \cite{vogels2011inhibitory,luz2012balancing}. Thus, it remains unclear whether STDP can account for the distribution of the preferred phases of L2/3 neurons.

Here we examined the hypothesis that STDP may underlie the unexplained distribution of preferred phases in L2/3. We analyzed under which conditions inhibitory STDP of L4I-to-L2/3 synapses can generate a non-trivial distribution of preferred phases, and investigated how the STDP rule shapes the resultant distribution. 
Below, we first define the network architecture and the response statistics of L4I neurons to whisking, in \cref{Architecture}. The STDP model is defined in \cref{STDPmodel}. Next, in \cref{single} we study the STDP dynamics of a single synapse in the limit of weak coupling. This will serve as a `free theory' that will later shed light on the STDP dynamics of a population of synapses. Then, in \cref{model} we analyze the STDP dynamics in the special case of an isotropic L4I population, i.e., with a uniform distribution of preferred phases, $\kappa_{ \mathrm{L4I}} = 0$. Subsequently, we turn to investigate the effect of $\kappa_{ \mathrm{L4I}} > 0$ and study how the STDP rule shapes the resultant distribution of phases in L2/3. Finally, we summarize our results, discuss the limitations of this study, and propose several empirical predictions to test our hypothesis.

\section{\label{Results}Results}

\subsection{\label{Architecture} Network model and order parameters}
We study STDP dynamics in a purely feed-forward architecture \hyperlink{fig 1}{Fig.\ 1(g)}. Each downstream L2/3 neuron receives input from $N$ L4I whisking neurons and a constant excitatory input. The excitatory input is assumed to be non-rhythmic and independent of the whisking phase. To facilitate the analysis, the downstream neurons are assumed to be delayed linear neurons following \cref{eq:postfiring3}. 

The spiking activity of the L4I whisking neurons is modeled by independent inhomogeneous Poisson processes with mean instantaneous firing rates (intensity), given by
\begin{equation}
    \label{eq:meanfiring4}
    \langle \rho_{i} (t) \rangle=D (1+\gamma \cos[ \theta(t)-\phi_{i}]),
\end{equation} 
where $ \rho_{i}(t)=\sum_{k} \delta(t-t_{i,k})$ is the spike train of neuron $i$ with $t_{i,k}$ denoting the time of the $k$th spike of the L4I neuron $i$ with preferred phase $\phi_i$. Parameter $D$ is the mean firing rate during whisking (in the air, averaged over one cycle), and $\gamma$ is the modulation depth. The phase along the whisking cycle is taken to be $\theta(t) = \nu t$ with $\bar{\nu} = \nu/(2 \pi) = 1/T_w$ denoting the whisking frequency. The angular brackets $\langle \cdots \rangle$ denote averaging with respect to the statistics of the noisy neuronal responses.

Under these assumptions, the firing rate of the downstream neuron will also oscillate at the whisking frequency, $\bar{\nu}$, with
\begin{equation}\label{eq:meanfiring4single}
    \langle \rho_{\text{post}} \rangle=  D_{\text{post}} (1+  \gamma_{\text{post}}  \cos[ \nu t  -\phi_{\text{post}}])
\end{equation} 
where from \cref{eq:postfiring3,eq:meanfiring4} the mean firing rate, amplitude modulation and preferred phase of the downstream neuron are determined by global order parameters that characterize the synaptic weights profile:
\begin{align}
    \label{eq:w_bar}
    & \bar{w} = \frac{1}{N}\sum_{k=1}^{N}w_k
    \\
    \label{eq:w_tilde}
    & \tilde{w} e^{i \psi} = \frac{1}{N}\sum_{k=1}^{N}w_k e^{ i\phi_k}
\end{align}
where $\bar{w} $ is the mean synaptic weight, and $ \tilde{w} e^{i \psi}$ is its `population vector' with $ \mathbb{R} \ni \tilde{w} \geq 0$. Thus yielding:
\begin{align}
\nonumber
    & D_{\text{post}} = I_{ex} - D \bar{w}, 
    \\
    \label{eq:PostStat}
    & \gamma_{\text{post}} = D \gamma \tilde{w} / D_{\text{post}}, 
    \ \ \ \mathrm{and}
    \\  
\nonumber
    & \phi_{\text{post}} = \pi + \psi + d \nu.
\end{align}

\subsection{\label{STDPmodel} The STDP model}
STDP can be thought of as a microscopic unsupervised learning rule in which the synaptic weight is modified according to the temporal relation of pre- and post-spike times. A wide range of STDP rules has been observed empirically \cite{markram1997regulation,bi1998synaptic,sjostrom2001rate,zhang1998critical,Abbott2000,froemke2006contribution,Nishiyama2000CalciumSR,shouval2002unified,WOODIN2003807}. For example, Bi and Poo reported a temporally asymmetric STPD rule in which an excitatory synapse was potentiated (strengthened) when the firing was causal, i.e., the pre-synaptic neuron fired about $20ms$ before post, and depressed (weakened) when the firing was acausal \cite{bi1998synaptic}. The STDP of inhibitory synapses has also been reported. Woodin and colleagues described a temporally symmetric rule \cite{WOODIN2003807}, whereas Haass and colleagues reported (in a different brain region) a temporally asymmetric STDP rule for inhibitory synapses \cite{Haas2003}. 

We write the STDP rule as a sum of two processes, potentiation, and depression (see also \cite{luz2012balancing,gutig2003learning,morrison2008phenomenological,OckerDoiron2015,luz2016oscillations,sherf2020multiplexing,sherfshamir2021}:
\begin{equation}
\label{eq:deltaw4}
    \Delta{w}=\lambda[f_{+}(w)K_{+}(\Delta t)-f_{-}(w)K_{-}(\Delta t)],
\end{equation}
where $\Delta w$ is the change in the synaptic weight $w$, $\lambda$ is the learning rate, and $\Delta t = t_{ \mathrm{post}} - t_{ \mathrm{pre}}$  is the time difference between the pre- and post-synaptic spike times. The first term on the right-hand side of \cref{eq:deltaw4} represents the potentiation ($+$) and the second term the depression ($-$). For mathematical convenience, we assumed a separation of variables, and thus write each process (potentiation and depression) as a product of a weight-dependent function $f_{\pm}(w)$ and a temporal kernel $K_{\pm}(\Delta t)$.

As in G\"{u}tig et al \cite{gutig2003learning}, we used the following choice for the synaptic dependence for the STDP rule 
\begin{align}
\label{eq:fplusminus4}
    f_{+}(w)&=(1-w)^{\mu} \\
    f_{-}(w)&= w^{\mu},
\end{align}
where $\mu \in [0, 1]$ controls the non-linearity of the learning rule.

In our numerical analysis, we used Gaussian kernels for the temporal dependence of the STDP rule:
\begin{equation}\label{eq:kernelSymmetric4}
    K_{\pm}(\Delta t)=\frac{1}{\tau_\pm \sqrt{2 \pi}} e^{-\frac{1}{2} (\frac{\Delta t-T_\pm}{\tau_\pm})^2},
\end{equation} 
where $\tau_\pm$ and $T_\pm$ are the widths and centers of the temporal kernels, respectively. 
Specifically, we focus on two families of learning rules. One, as in  Woodin et al \cite{WOODIN2003807}, is a temporally symmetric difference of Gaussians `Mexican hat' learning rule, in which $T_+ = T_- = 0$. We shall term the upright Mexican hat rule, $\tau_+ < \tau_-$, Hebbian, and the inverted Mexican hat rule, $\tau_+ > \tau_-$, anti-Hebbian. 

The other, as in Haas and colleagues \cite{haas2006spike}, is a temporally asymmetric rule with $T_{\pm}  \neq 0$. We shall term the (anti-) Hebbian the case of ($T_+ <0$ AND $T_- > 0$) $T_+ >0$ AND $T_- < 0$. In the limit of $\tau_{\pm} \rightarrow 0$ the temporal kernels converge to a Dirac delta function, $K_{\pm}(\Delta t) \rightarrow \delta (\Delta t - T_{\pm})$.

In the limit of slow learning, $\lambda \rightarrow 0$, STDP dynamics effectively averages the noise in the neuronal activity. The fluctuations in the synaptic weights become negligible and the stochastic dynamics of the synaptic weights can then be replaced with deterministic dynamics for their means (see \cite{gutig2003learning,gilson2012frequency,luz2014effect,luz2016oscillations,sherf2020multiplexing,sherfshamir2021}, yielding
\begin{equation}\label{eq:wdot4}
    \frac{\dot{w}_{j}(t)}{\lambda}=I^+_{j}(t) -I^-_{j}(t) 
\end{equation}
where
\begin{equation}\label{eq:wdotpm4}
    I^\pm_{j}(t)=f_{\pm}(w_{j}(t)) \int_{-\infty}^{\infty}\Gamma_{j, \text{L2/3}}( \Delta)K_{\pm}(\Delta)d\Delta,
\end{equation}
and 
\begin{equation}\label{eq:CorrDef}
    \Gamma_{j, \ \text{L2/3}}(\Delta) =
    \int_0^{T_w} \frac{dt}{T_w} \langle \rho_j(t) \rho_{ \text{L2/3}} (t + \Delta) \rangle 
\end{equation}
is the temporal average of the cross-correlation between the $j$th L4I pre-synaptic neuron and the L2/3 post-synaptic neuron, averaged over one period of the whisking cycle, $T_w$ 
see \cref{correlderivation}.

\subsection{\label{single}STDP dynamics of a single synapse}
We begin by analyzing the simple case in which only a single L4I-to-L2/3 synapse is plastic. Although highly artificial, this case constitutes a `free theory' that will prove pivotal later. In the limit of large $N$, the activity of the downstream neuron will be almost unaffected by the activity of the single plastic neuron. In this case, the firing rate of the downstream neuron will oscillate at the whisking frequency, $\bar{\nu}$, with a fixed preferred phase, $\phi_{\text{post}} $, determined by its non-plastic inputs, following \cref{eq:meanfiring4single,eq:PostStat}. 

In the limit of weak coupling, the cross-correlation between the plastic pre-synaptic neuron and the post-synaptic cell can be approximated by the product of their mean responses, yielding for large $N$ 
\begin{equation} \label{eq:SingleCellCorr}
    \Gamma_{\text{pre}, \ \text{L2/3}}(\Delta) = 
    D_{\text{pre}} D_{\text{post}}
    \left(
    1 + \frac{\gamma_{\text{pre}} \gamma_{\text{post}}}{2} 
    \cos[ \phi + \Delta \nu ]
    \right)
\end{equation}
where $ \phi = \phi_{\text{pre}} - \phi_{\text{post}} = \phi_{\text{pre}} -  (\pi + \psi + d \nu) $ is the phase difference between the pre- and post-post synaptic neurons. Substituting \cref{eq:SingleCellCorr} into \cref{eq:wdot4,eq:wdotpm4} one obtains that the synaptic weight converges to a single fixed point 
\begin{align}
    \label{eq:fpsinglesol}
    & w^*( \phi)  =  \frac{1}{\big(
    Q(\phi)\big)^{1/\mu}+1}
    \\
    & Q( \phi)  =  \frac{ 
    \bar{K}_{-}+\eta \tilde{K}_- \cos(\Omega_- - \phi) }{
    \bar{K}_{+}+\eta \tilde{K}_+ \cos(\Omega_+ - \phi)}
\end{align}
where $\eta = \frac{\gamma_{\text{pre}} \gamma_{\text{post}}}{2} $, and $\bar{K}_{\pm}$ and $\tilde{K}_{\pm} e^{i \Omega_{\pm}}$ are the Fourier transforms of the STDP kernels:
\begin{align}\label{eq:Kfourier4}
    &\bar{K}_{\pm} = \int_{-\infty}^{\infty} K_{\pm}(\Delta) d\Delta\\
    & \label{eq:Kfourier5}
    \tilde{K}_{\pm}  e^{i \Omega_{\pm}} = \int_{-\infty}^{\infty} K_{\pm}(\Delta) e^{-i  \nu \Delta} d\Delta
    .
\end{align}
Note that in our specific choice of kernels, $\bar{K}_{\pm}=1$, by construction.

Part of the utility of this simplified scenario is that it enables a complete understanding of how different parameters affect the resultant fixed-point, \cref{eq:fpsinglesol}. In the limit of weak coupling, the STDP dynamics of a single inhibitory synapse is similar to that of an excitatory synapse, which was analyzed in \cite{luz2016oscillations}. In brief, the synapse is potentiated or depressed depending on the phase difference between the pre- and the post-synaptic neurons, $\phi$. The parameter $\mu$ governs the smoothness of the transition between potentiated and depressed synapses. In the additive rule, i.e., $\mu = 0$, the synapses are either potentiated to 1 or depressed to 0 with a sharp transition. As the value of $\mu$ increases, the transition becomes smoother, \hyperlink{Fig2}{Figs.\ 2(a)-2(b)}.

The temporal structure of the STDP rule determines which phases will be potentiated and which will be depressed. 
It is convenient to define the width of the profile as the range of values of $\phi$ such that $w^* ( \phi) \geq 1/2$, and critical values of $ \phi$ by the condition $Q ( \phi_c) = 1$, yielding
\begin{equation}\label{eq:muinvariant}
    \phi_{c_{1,2} } = \alpha_0 + \psi \pm \pi/2
\end{equation}
with the definitions 
\begin{align} \label{eq:alpha0}
    & \tilde{K} e^{ i \alpha_0} = 
    \tilde{K}_- e^{i (\Omega_- + \nu d)} - 
    \tilde{K}_+ e^{i (\Omega_+ + \nu d)}  
    .
\end{align}
Thus, the width of the synaptic weights profile is always $\pi$ \footnote{The width can be governed by adding a parameter that controls the relative strength of depression [CITE LuzShamir16]}. For a temporally symmetric learning rule $\Omega_\pm = 0$, yielding a symmetric profile, $\phi_{ \mathrm{post}} - \phi_c =\pm \pi /2$ .
A Hebbian learning rule will potentiate synapses with a phase difference that is small in absolute values, $ | \phi | < \pi/2$. Whereas, an anti-Hebbian temporally symmetric rule will potentiate synapses with $ | \phi | > \pi/2$, \hyperlink{Fig2}{Fig.\ 2(a)}. 

For the temporally asymmetric delta rule, in the limit of $\tau_\pm \rightarrow 0$, $\tilde{K}_\pm = 1$ and $\Omega_\pm = \nu T_\pm$, yielding $\phi_{ \mathrm{post}} - \phi_{c1} = \nu \frac{T_+ + T_-}{2} $ and $\phi_{ \mathrm{post}} -\phi_{c2} = \nu \frac{T_+ + T_-}{2} +\pi $.  The resultant profile for the Hebbian rule ($T_+ >0$ and $T_- <0$) is shown in \hyperlink{Fig2}{Fig.\ 2(b)}.



\begin{figure*} \hypertarget{Fig2}{}
	\centering
		\begin{subfigure}[t]{0.01\textwidth} \vspace{-3.5cm}
		\textbf{(a)}
	\end{subfigure}
	\begin{subfigure}[t]{0.3\textwidth}
		\centering
		\includegraphics[width=\linewidth]{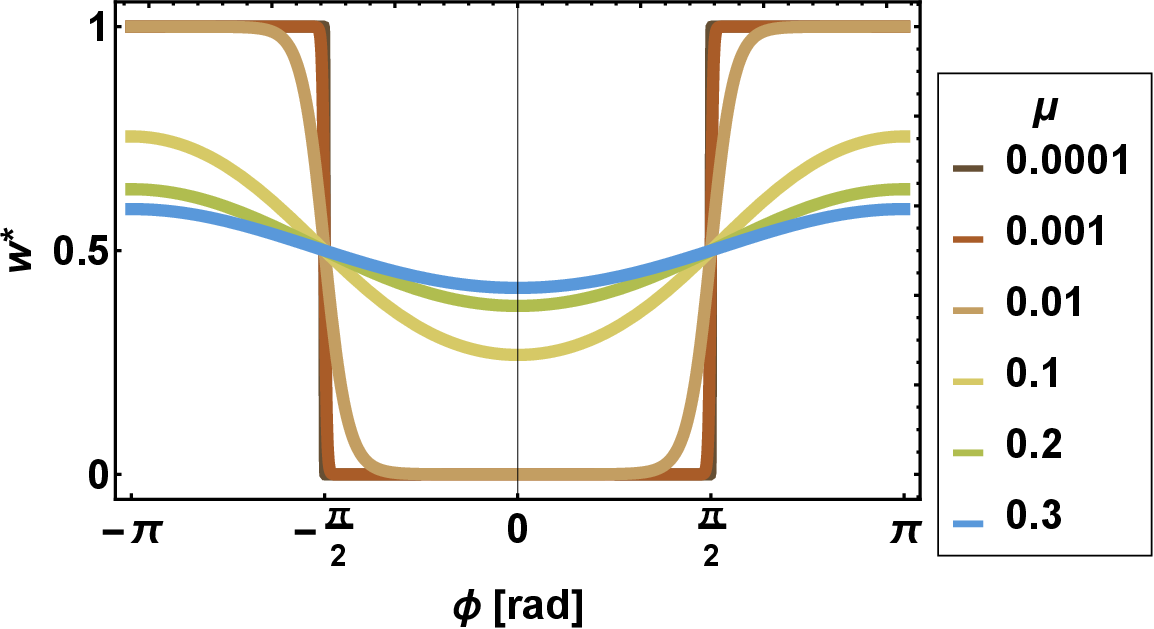} 
	\end{subfigure}
	 \hspace{0.1cm}
		\begin{subfigure}[t]{0.01\textwidth} \vspace{-3.5cm}
		\textbf{(b)} 
	\end{subfigure}
	\begin{subfigure}[t]{0.3\textwidth}
		\centering
		\includegraphics[width=\linewidth]{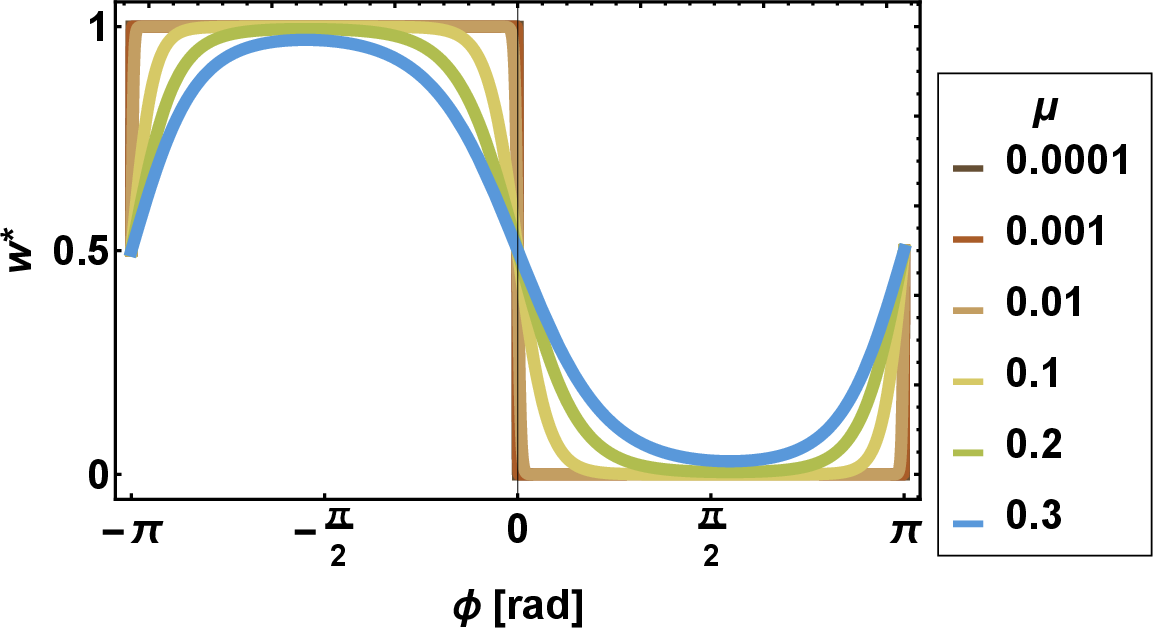} 
	\end{subfigure}

		\vspace{1cm} 	 
		\begin{subfigure}[t]{0.01\textwidth} \vspace{-3.5cm}
		\textbf{(c)} 
	\end{subfigure}
	\begin{subfigure}[t]{0.3\textwidth}
	\centering
	\includegraphics[width=\linewidth]{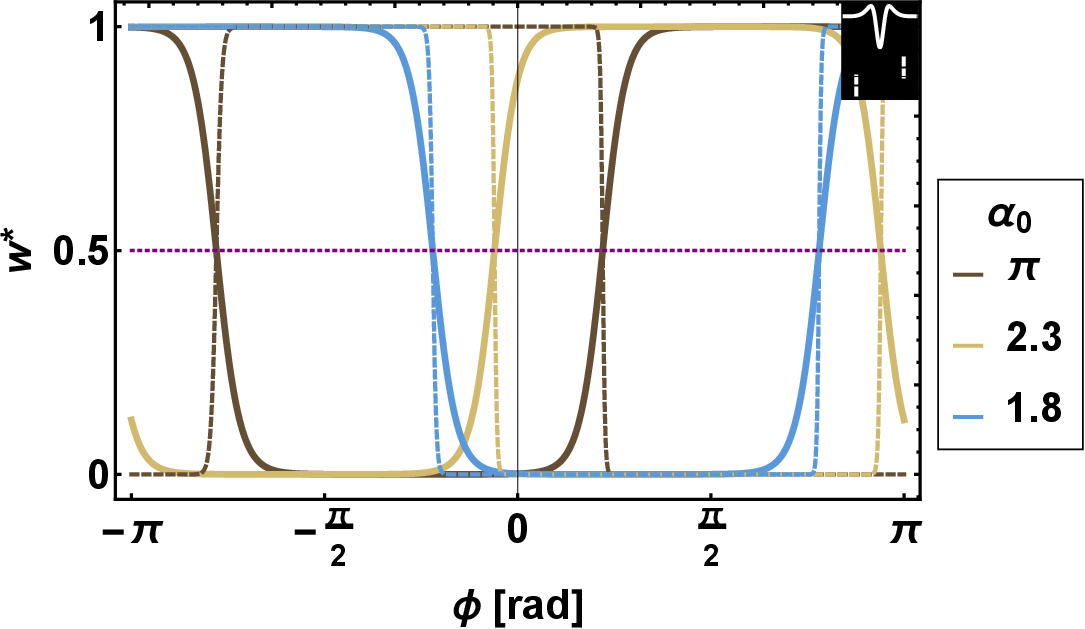} 
\end{subfigure}
	 \hspace{0.1cm}
	\begin{subfigure}[t]{0.01\textwidth} \vspace{-3.5cm}
	\textbf{(d)} 
\end{subfigure}
\begin{subfigure}[t]{0.3\textwidth}
	\centering
	\includegraphics[width=\linewidth]{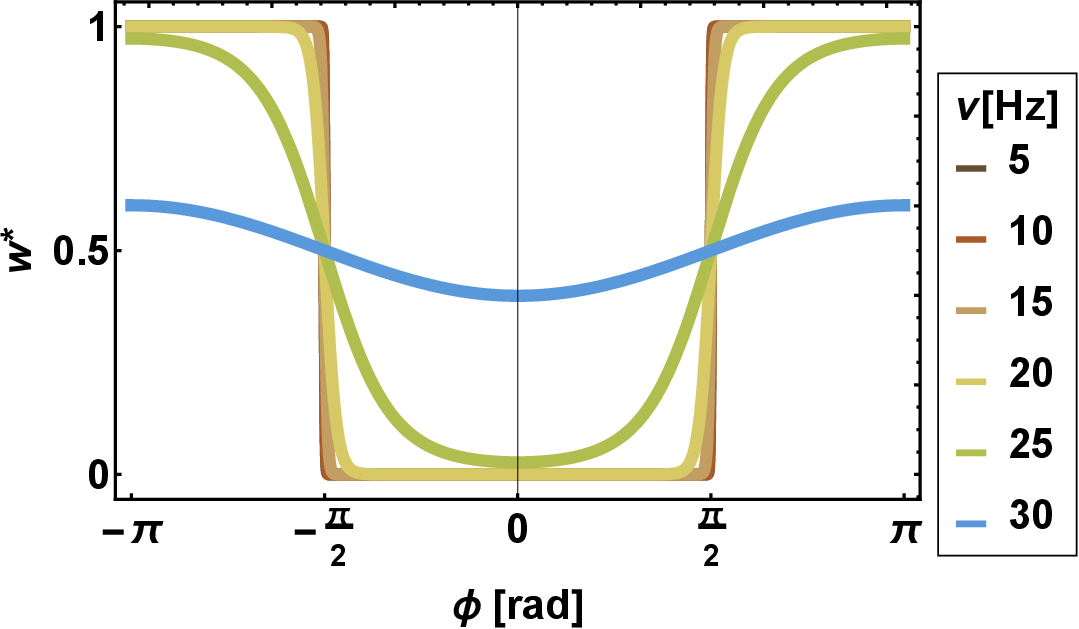} 
\end{subfigure}
  \hypertarget{Fig4.5}{}
	\caption{\textbf{Fixed-point solutions for STDP dynamics of a single synapse}.
	The different panels show $w^*$ as a function of $\phi$, see \cref{eq:fpsinglesol}, for different sets of parameters (depicted by colors). (a) For different values of $\mu$ for the symmetric kernel, \cref{eq:kernelSymmetric4}. (b) For different values of $\mu$ for the delta kernel. (c)   For different values of $\alpha_0$ for both the symmetric (solid lines) and the delta (dashed lines) rules. The dashed horizontal purple line depicts  $w^*(\phi_c)=0.5$. (d) For different values of $\nu$ for the symmetric rule.
Unless stated otherwise, the parameters for the symmetric kernel are $\tau_-=20\text{ms}$, $\tau_+=50\text{ms}$, $\mu=0.01$, $\nu=14 \text{hz}$ and  $\Gamma_1=0.5$. The parameters for the delta kernel are  $T_-=-20\text{ms}$, $T_+=20\text{ms}$, $\mu=0.01$, and $\nu=14 \text{hz}$. In panel (c) the symmetric kernel was plotted with $T_+ \approx 20, 30, -10 \text{ms}$ and the delta kernel was plotted with $T_-\approx 35, 15, 4 \text{ms}$, yielding $\alpha_0 = \pi, 2.3, 1.8$, respectively.}
\end{figure*}

\subsection{\label{model}STDP in an isotropic model}
We now turn to study the STDP dynamics of a population of $N$ L4I-to-L2/3 inhibitory synapses, \hyperlink{fig 1}{Fig.\ 1(g)}. First, we consider the case of an isotropic population, $\kappa_{ \text{L4I}} = 0$, taking the preferred phases of the L4I neurons to be evenly spaced on the ring, $\phi_k = 2 \pi k /N$. The biological scenario of $\kappa_{ \text{L4I}} > 0$ is addressed in \cref{kappaNonZero}.

\subsubsection{\label{uniform}The uniform fixed point}
In the isotropic case, $\kappa_{ \text{L4I}} = 0$, a uniform solution, where $w_k = w^*, \ \forall k \in \{1 \ldots N \} $ always exists. In the uniform solution $\bar{w} = w^*$ and $\tilde{w} = 0$. Consequently, $\gamma_{\text{post}} = 0$ and the whisking signal is not transmitted downstream. Thus, the uniform solution prevents the transmission of the whisking signal. Below we study the stability of the uniform solution in the limit of large $N$; see \cref{fpsstability} for details. In this limit, there are two types of uniform fixed-point solutions:
\begin{align}\label{eq:steadystate4}
    & w^*_1 = \frac{1}{2} & (type \ 1) \\
    \label{eq:steadystate5}
    & w^*_2 = \frac{I_{ex}}{D} 
    & (type \ 2)
\end{align}
In \emph{type 1}, potentiation and depression cancel each other out via the weight-dependence of the STDP rule, $f_+ (w^*_1) = f_-(w^*_1)$. In \emph{type 2}, the mean inhibitory input to the downstream neuron balances the excitatory input, $I_{ex} - D w^*_2 = 0$.

Along the uniform direction, the stable fixed point is given by $\min \{ w^*_1, w^*_2\}$. To further study the stability of the fixed points we expand the dynamics to first order in fluctuations $\delta w(\phi_j) = w(\phi_j) - w^*$, yielding  
\begin{equation}\label{eq:stabilityM}
    \frac{\delta \mathbf{\dot{w}}(\phi)}{\lambda}=\mathbf{M} \delta \mathbf{w}.
\end{equation}
where $\mathbf{M}$ is the stability matrix. 

The stability matrix has two prominent eigenvalues. One, $m_u$, is in the uniform direction, $\bar{w}$. The other, $m_w$, is in the `whisking' direction, $\tilde{w}$ (with degeneracy 2). The eigenvalues of the stability matrix around $w^*_1$ in this limit were calculated and yielded 
\begin{align}
    \label{eq:lambdabar1}
    & m_{u} =     - \mu D^2 \Big( \frac{ I_{ex} }{D} - \frac{1}{2} \Big) 2^{2-\mu}
    \\
    \label{eq:lambdatilde1}
    & m_{w} \approx m_{u,1} +\frac{1}{2^{2 + \mu }} D^2 \gamma^2   
    \tilde{K} \cos( \alpha_0 ) .
\end{align}
Thus, the uniform fixed point $w^*_1$ is stable with respect to fluctuations in the uniform direction, $m_{u} <0 $, for $ I_{ex} > D/2$. In the type 1 fixed point, $m_{u}$ provides a stabilizing term against fluctuations in the whisking direction, $m_{w}$, \cref{eq:lambdatilde1}. This stabilizing term can become arbitrarily small by decreasing $\mu$; thus enabling fluctuations in the whisking direction to develop and destabilize the homogeneous fixed point. For $ I_{ex} > D/2$ there is a critical value of $\mu$, 
\begin{align} \label{eq:mu_crit}
    & \mu_{\mathrm{crit}} = 
    \frac{16 \gamma^2}{ I_{ex} / D - 1/2}  \tilde{K} \cos(\alpha_0)
\end{align}
such that for any $\mu < \mu_{\mathrm{crit}}$, the uniform solution is unstable in the whisking direction. Since $\mu \geq 0$, a critical value exists if and only if  $ \cos( \alpha_0)  > 0$.   

An analysis of the stability of type 2 (balanced) uniform solution, $w^*_2$, in the limit of large $N$, indicates that the two prominent eigenvalues are given by  
\begin{align}\label{eq:lambdabar2}
    m_u =  & D^2\big( (w^*_2)^\mu -  (1-w^*_2)^\mu \big), 
    \\
    \nonumber
    m_w = & \frac{1}{4} D^2 \gamma^2  \big( f_-(w_2^*) \tilde{K}_-  \cos(\nu d+\Omega_-) -\\ 
    & f_+(w_2^*) \tilde{K}_+ \cos(\nu d+\Omega_+) \big).
\end{align}	
Type 2 fixed-point is stable with respect to uniform fluctuations for $I_{ex} < D/2$. In contrast to the type 1 fixed-point, here $m_u$ does not provide a stabilizing term against fluctuations in other directions - as expected from the balanced solution; see \cite{luz2012balancing}. 

Similar to type 1 fixed-point, the balanced fixed-point tends to lose stability in the whisking direction as $ \cos( \alpha_0) $ increases. For both types of uniform fixed points, in the limit of the additive learning rule, $\mu \rightarrow 0$, the uniform fixed point is unstable for $ \cos( \alpha_0)  > 0$. Thus, the temporal structure of the learning rule governs the stability of the uniform fixed point via a single parameter, $ \alpha_0$, as defined in \cref{eq:alpha0}.

For the temporally symmetric difference of Gaussians rule (\cref{eq:kernelSymmetric4} with $T_\pm = 0$) one obtains $  \tilde{K} e^{ i\alpha_0}  = \big( e^{- \frac{1}{2} ( \nu \tau_+)^2 } - e^{- \frac{1}{2} ( \nu \tau_-)^2 } \big) e^{ id \nu}$. In the case of a Hebbian (upright) Mexican hat ($\tau_+ < \tau_-$) one obtains $ \alpha_0 = d \nu$, whereas for the anti-Hebbian inverted Mexican hat ($\tau_+ > \tau_-$),   $ \alpha_0 = \pi + d \nu$. Thus, for short to moderate delays, $ d < \frac{1}{4} T_w$, the uniform fixed point will lose its stability for small $\mu$ in the anti-Hebbian rule.

For the temporally asymmetric rule, we consider the limit of $\tau_\pm \rightarrow 0$, in which the STDP kernels, $K_\pm(t)$ converge to a Dirac delta. In this case $ \tilde{K} \cos( \alpha_0)  =  \cos[\nu (d + T_-)] - \cos[\nu (d + T_+)] = 2 \sin \big( | \nu | [ \frac{T_+ + T_-}{2} + d ] \big) \sin( |\nu | \frac{T_- - T_+}{2})$. Thus, for example, in a Hebbian temporally asymmetric rule $T_- <0$ and $T_+ >0$, for $|T_-| +|T_+| < T_w$, $ \cos( \alpha_0) $ is positive for $ |T_- | - 2d < T_+ < T_w + |T_- | - 2d  $.

Thus overall, the uniform solution does not allow for the transmission of the whisking signal downstream. However, for a wide range of parameters, this solution is unstable. In this case, one expects that a non-uniform synaptic weights profile will develop and facilitate the transmission of the whisking signal downstream.

\subsubsection{\label{IsotropicLimitCycle}The limit cycle solution}

\hyperlink{Fig3}{Figure\ 3} shows the STDP dynamics when the uniform fixed-point is unstable. As shown in the figure, the synaptic weights do not converge to a non-uniform fixed point. Rather, the dynamics converge to a limit cycle, in which all the synaptic weights vary across their entire dynamic range, \hyperlink{Fig3}{Figs.\ 3(a)-3(c)}. 
Nevertheless, the whisking signal is transmitted downstream, and its power, $ D_\mathrm{post} \gamma_\mathrm{post}$, remains steady. This results from the fact that the global order parameters $\bar{w}$ and $\tilde{w}$ (see  \cref{eq:w_bar,eq:w_tilde}) converge to a stable fixed-point, \hyperlink{Fig3}{Fig.\ 3(d)}. In contrast, the phase of the downstream neuron, $\phi_\mathrm{post}  = \psi(t) + \nu d - \pi$ drifts in time with a constant drift velocity $ v_\mathrm{drift}$, \hyperlink{Fig3}{Fig.\ 3(e)}. 

Can a non-uniform fixed-point exist? Let us assume that the synaptic weights converged to a non-uniform fixed-point. In this case, the downstream neuron would fire rhythmically at a fixed rate, modulation depth and phase according to \cref{eq:meanfiring4single,eq:PostStat}. Hence, the system is under the terms of section STDP for a single synapse, as detailed in \cref{single}. As a result, the synaptic weights must obey the `free theory', \cref{eq:fpsinglesol}, that determines the order parameters: $\bar{w}$, $\tilde{w}$, and $\psi$. The order parameters of the `free theory' dictate the activity of the downstream neuron, via $D_{\text{post}}$, $\gamma_{\text{post}}$, and $\phi_{\text{post}}$ (see \cref{eq:PostStat}), which, in turn, must be consistent with the initially assumed activity.

This self-consistency argument is illustrated in \hyperlink{Fig3}{Figs.\ 3(g)-3(i)}. 
Assuming the system converged to a non-uniform fixed point and that w.l.g.\ $\phi_\mathrm{post}^\mathrm{(ass)} = 0$, the synaptic weight profile, given by \cref{eq:fpsinglesol}, is depicted by the black line. The phase of the rhythmic input to the downstream neuron, $\psi + \pi$, is shown by the dashed red line. Consequently, the preferred phase of the downstream neuron is given by $\phi_\mathrm{post}^\mathrm{(s.c.)}  = \psi + \nu d + \pi$ (green dashed line). In \hyperlink{Fig3}{Fig.\ 3(g)} $\phi_\mathrm{post}^\mathrm{(s.c.)} - \phi_\mathrm{post}^\mathrm{(ass)} = 0$, and the self-consistency condition is satisfied. 

In contrast, in \hyperlink{Fig3}{Figs.\ 3(h)-3(i)} the self-consistency condition is not satisfied. In \hyperlink{Fig3}{Fig.\ 3(h)}  $\phi_\mathrm{post}^\mathrm{(s.c.)} - \phi_\mathrm{post}^\mathrm{(ass)} > 0 $, which induces a positive drift velocity, as shown by the blue square in \hyperlink{Fig3}{Fig.\ 3(f)}.  Whereas in \hyperlink{Fig3}{Fig.\ 3(i)}  $\phi_\mathrm{post}^\mathrm{(s.c.)} - \phi_\mathrm{post}^\mathrm{(ass)} < 0 $, induces a negative drift velocity, as shown by the green square in \hyperlink{Fig3}{Fig.\ 3(f)}.

Thus, a non-uniform fixed-point (for $\kappa_{\mathrm{L4I}} = 0$) is a special case in which $ v_\mathrm{drift} = 0$. In the generic case $ v_\mathrm{drift} \neq 0$. We find that, for small $\mu$, the drift velocity can be approximated by (see \cref{phasedynkappa}):
\begin{align}\label{eq:Vdrift0}
    v_\mathrm{drift} = \frac{\lambda }{4} D^2 \gamma^2 \tilde{K} \left(
    3 \alpha_0 \sin (\alpha_0)
    + \cos(2 \alpha_0)-\cos(\alpha_0)
    \right),
\end{align}
compare the solid line, \cref{eq:Vdrift0}, and open squares (numerical) in \hyperlink{Fig3}{Fig.\ 3(f)}. Thus, the drift velocity is determined by the STDP rule; however, only via $\tilde{K}$ and $ \alpha_0$. In particular, the sign of $ v_\mathrm{drift} $ is determined solely by $\alpha_0$ in accordance with the self-consistency argument above.

\begin{figure*}[htp] \hypertarget{Fig3}{}
\centering
	\begin{subfigure}[t]{0.01\textwidth} \vspace{-4.3cm}
		\textbf{(a)}
	\end{subfigure}
\includegraphics[width=.3\linewidth]{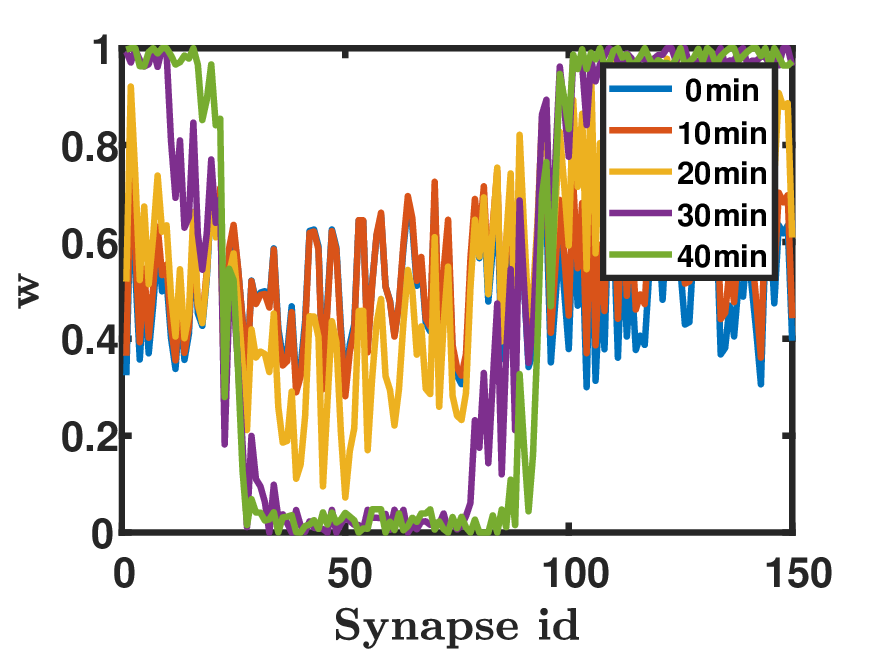}\quad
	\begin{subfigure}[t]{0.01\textwidth} \vspace{-4.3cm}
		\textbf{(b)} 
	\end{subfigure}
\includegraphics[width=.3\linewidth]{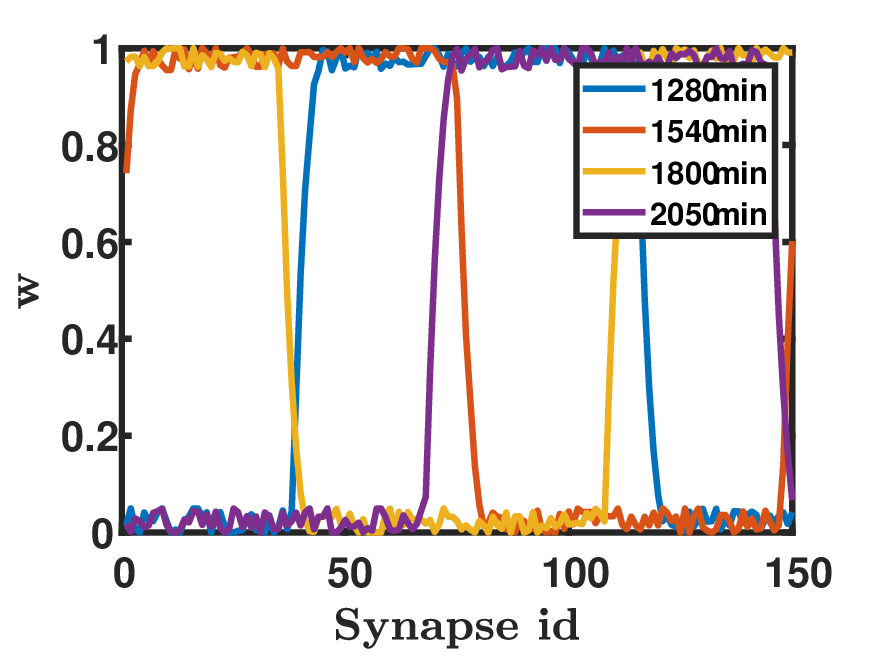}\quad
	\begin{subfigure}[t]{0.01\textwidth} \vspace{-4.3cm}
		\textbf{(f)} 
	\end{subfigure}
\includegraphics[width=.3\linewidth]{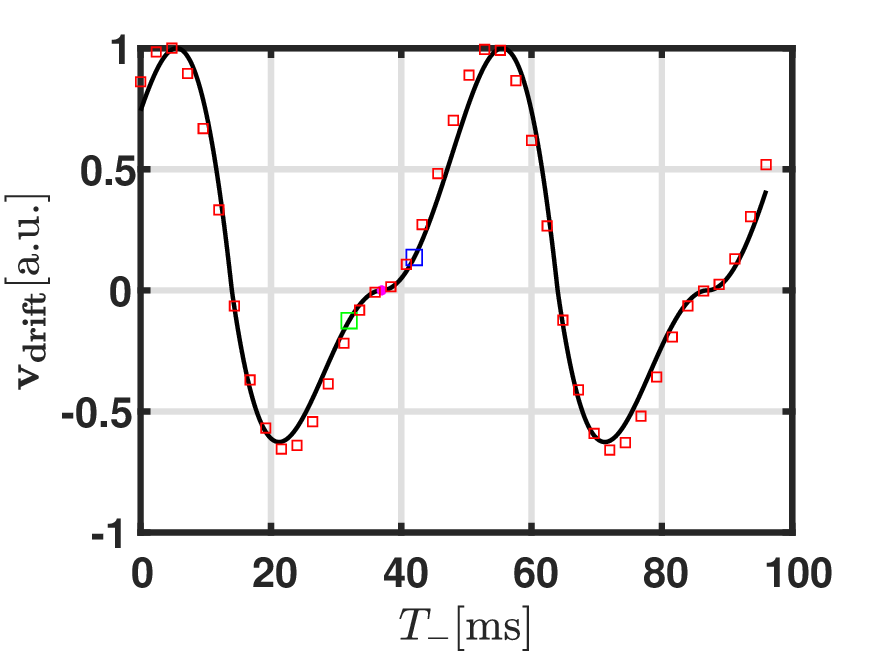}

\medskip

\includegraphics[width=.45\linewidth]{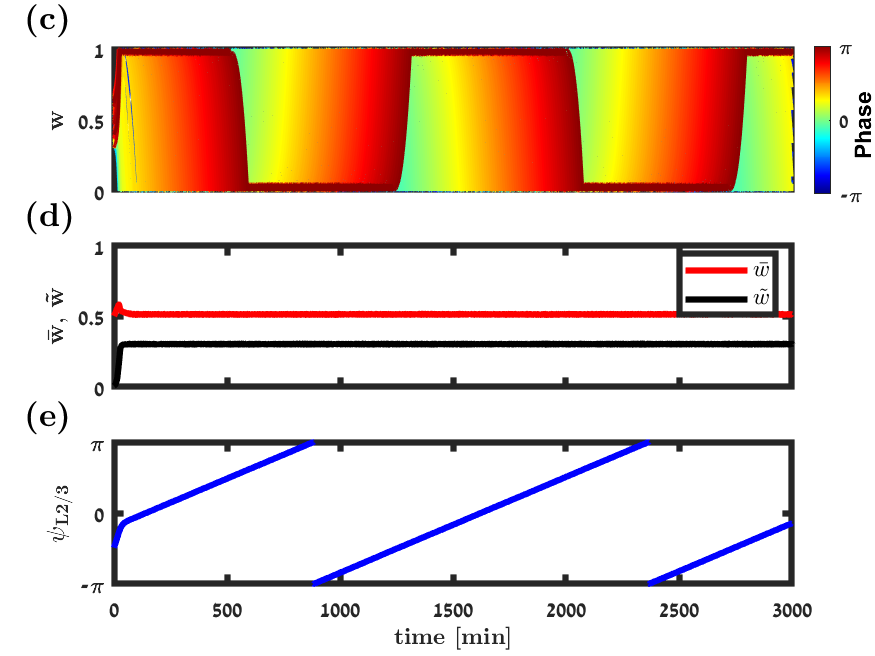}\quad
\includegraphics[width=.45\linewidth]{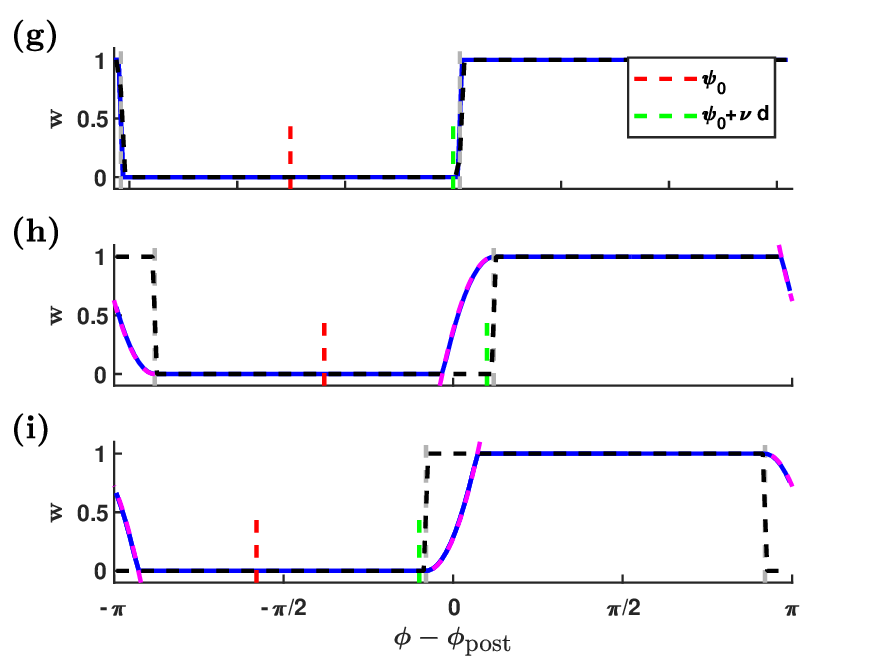}

\caption{\textbf{STDP dynamics for $\kappa=0$}.
 (a) and (b) The synaptic weights profile, $w(\phi,t)$, is shown as a function of the preferred phases of the upstream neurons, $\phi$, at different times (differentiated by color). (c) Trace depicts synaptic weights are shown as a function of time. The synaptic weights are differentiated by color according to the preferred phase of the upstream neuron. (d) and (e) Dynamics of the order parameters $\bar{w}$ (red) \& $\tilde{w}$ (black) are shown in d, and $\psi$ in e. (f) Drift velocity is shown as a function of $T_-$. The black line follows \cref{eq:Vdrift0}, and red squares depict the estimation of the drift velocity from the numerical simulations. (g)-(i) Synaptic weights profile and the self-consistency argument. The black lines depict the profile assuming zero-drift, i.e., according to \cref{eq:Vdrift0}. The vertical dashed grey lines (aligning with dashed black lines) show $\phi_{c_{1,2}}$. The input to the downstream neuron, $\psi_0+\pi$, and the phase of the downstream neuron, $\psi_0+\pi+\nu d$, assuming zero drift (i.e., the black profile), are shown by the dashed red and green lines, respectively.  
 The parabolic approximation of the semi-phenomenological model, \cref{eq:phenomenological}, is depicted by the dashed magenta lines. For comparison, the synaptic weights profile obtained by numerical simulation of the STDP dynamics is shown in blue as a function of $\phi-\phi_{post}$. 
The parameters used to generate the figure are as follows. In (a) - (e), the symmetric anti-Hebbian rule was used with random uniformly distributed initial conditions for synaptic weights in the interval $[0.3, 0.7]$, and   $\lambda=10^{-3}$, $D=10\text{hz}$, $I=10\text{hz}$, $\tau_-=20\text{ms}$, $\tau_+=50\text{ms}$,  $T_\pm=0\text{ms}$, $\mu=10^{-4}$, $\nu=7 \text{hz}$, $d=5 \text{ms}$ and  $\gamma=1$. In (f) - (i)  the asymmetric Delta rule was used with $\lambda=0.01$, $D=10\text{hz}$, $I=6\text{hz}$, $T_+=36\text{ms}$, $\mu=0$, $\nu=20 \text{hz}$, $d=12 \text{ms}$, $\gamma=1$, and $T_-=37, \ 42, \ 32\text{ms}$ in (g), (h), and (i), respectively. Initial conditions were $w(\phi, t=0)=0.5+0.3\cos(\phi)$. In all simulations, $N=150$ was used.}
		\label{fig:SingleSynapse}
\end{figure*}

\subsection{STDP dynamics for $\kappa_{\mathrm{L4I}} > 0$}\label{kappaNonZero}
\hyperlink{Fig4}{Figure 4(a)} shows the STDP dynamics in the case of an un-isotropic upstream population, $\kappa_{\mathrm{L4I}} > 0$. As in the isotropic case, for a wide range of parameters, the STDP dynamics converge to a limit cycle. In contrast with the isotropic case, the drift velocity is not constant and depends on the phase along the whisking cycle, \hyperlink{Fig4}{Figs.\ 4(b)-4(c)}. Consequently, the time the downstream neuron spends at each phase is not constant and is proportional to the inverse of the drift velocity in that phase, \hyperlink{Fig4}{Fig.\ 4(f)}. Thus, STDP dynamics induces a distribution over time for the preferred phase of single L2/3 neurons, which in turn is translated into a distribution over the downstream population, \hyperlink{Fig4}{Figs.\ 4(d)-4(e)}. 

For small $\mu$, a perturbation analysis for small $\kappa$ (see \cref{phasedynkappa}) yields
\begin{align}\label{eq:Vdrift1}
    v_\mathrm{drift} & =  v_\mathrm{drift}^{(0)} + 
    \kappa \lambda  D^2 \gamma^2 \tilde{K} F(\alpha_0) \sin (\psi) \\
    F(\alpha_0) & =  - \frac{3 \pi \alpha_0 \sin( \frac{3 \alpha_0}{2} )}{
    16 \tan( \frac{ \alpha_0}{2} )
    },
\end{align}
where $v_\mathrm{drift}^{(0)}$ is the drift velocity as $ \kappa = 0$, \cref{eq:Vdrift0}. For small $\kappa $, the drift velocity has a single minima, either at $\pi/2$ or at $-\pi/2$, depending on the STDP rule via the sign of $ F( \alpha_0 )$, see \cref{eq:Vdrift1}.

\hyperlink{Fig4}{Figure 4(h)} shows the color-coded distribution of preferred phases of the downstream population, for different values of $\kappa_{ \mathrm{L4I}}$. In the limit of an isotropic upstream population, $\kappa_{ \mathrm{L4I}} = 0$, the drift velocity is constant and the induced distribution of preferred phases in layer 2/3 will be uniform. For small $\kappa_{ \mathrm{L4I}} $, the peak of the distribution of the preferred phase in L2/3  will shift by either $\pi/2 +d\nu$ or $ - \pi/2 +d\nu$, relative to L4I, depending on the STDP rule. As $\kappa_{ \mathrm{L4I}}$ grows, the distribution of preferred phases in the upstream layer (L4I) converges to a delta function, the STDP dynamics will converge to a fixed-point and distribution of preferred phases in L2/3 will converge to a delta function at $d\nu$, as shown in inset \hyperlink{Fig4}{Fig .\ 4(h)}.

\begin{figure*}  \hypertarget{Fig4}{}
		\centering
		\begin{subfigure}[t]{0.45\textwidth} 
			\centering
			\includegraphics[width=\linewidth]{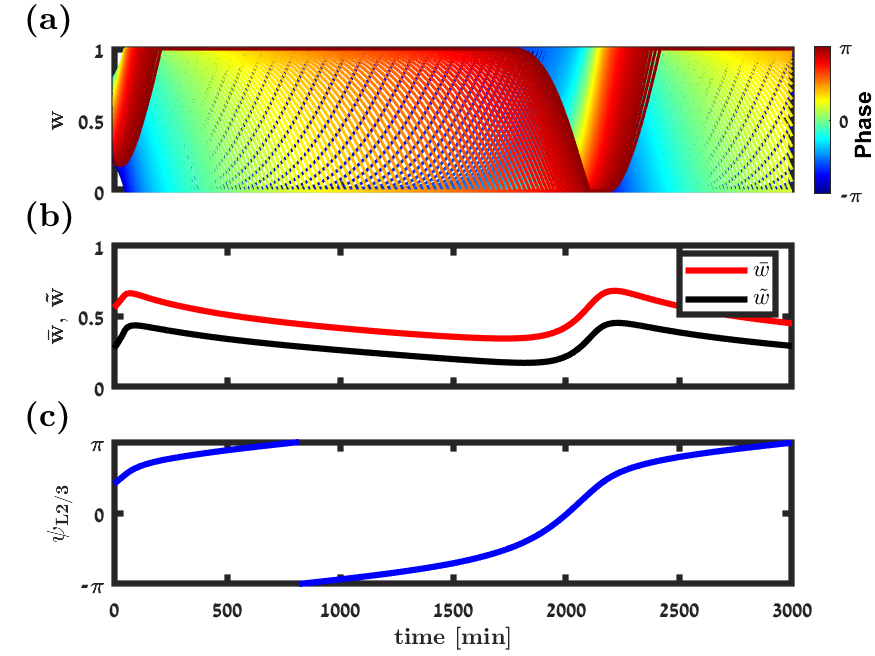} 
		\end{subfigure}
		\hspace{0.1cm}
		\begin{subfigure}[t]{0.5\textwidth}
			\centering
			\includegraphics[width=\linewidth]{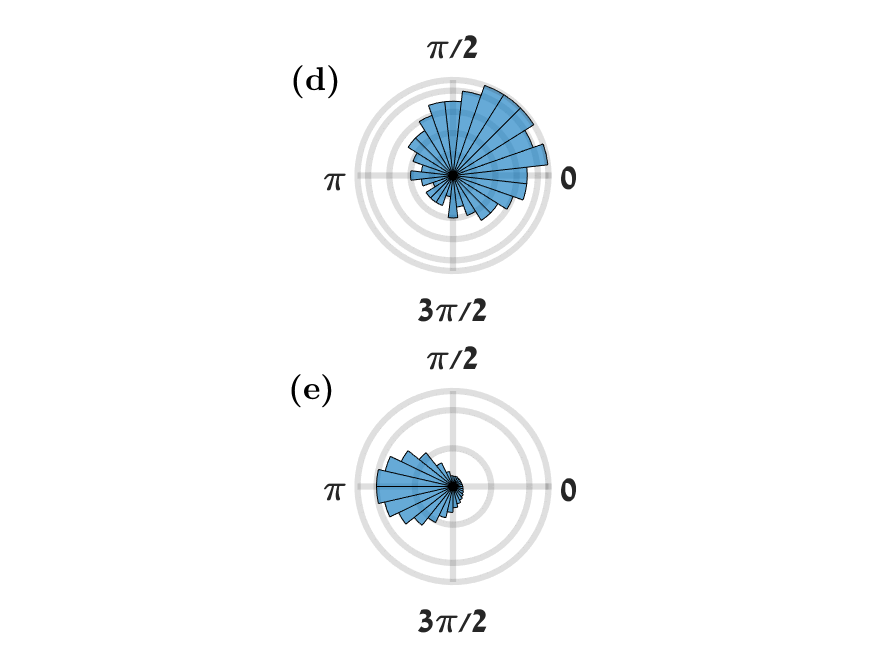} 
		\end{subfigure}\vspace{1cm}

	\begin{subfigure}[t]{0.45\textwidth}
	\centering
	\includegraphics[width=\linewidth]{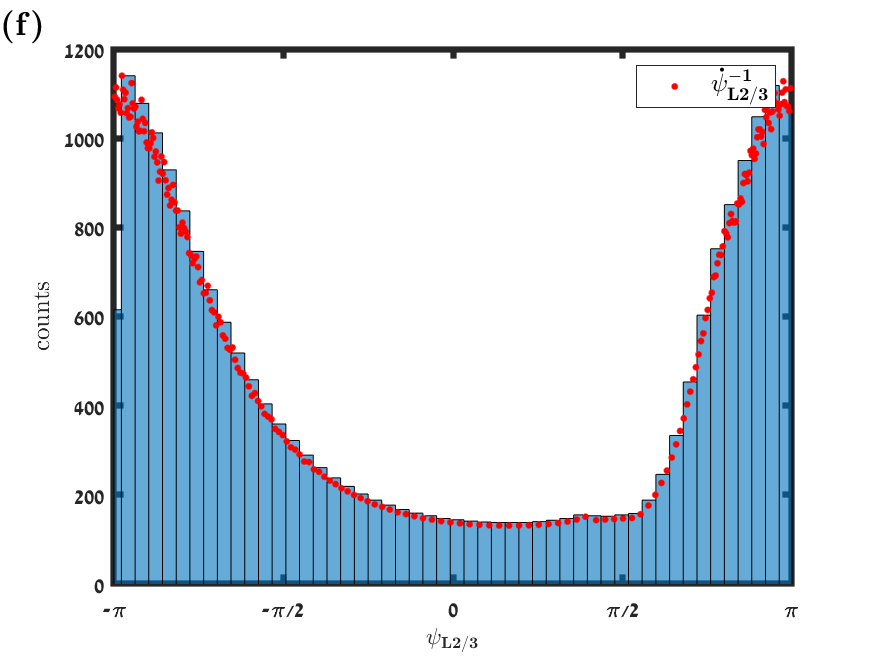} 
\end{subfigure}
\hspace{0.1cm}
\begin{subfigure}[t]{0.01\textwidth} \vspace{-6.0cm} \hspace{-0.4cm}
		\textbf{(g)}
	\end{subfigure}
		\begin{subfigure}[t]{0.45\textwidth}
			\centering
			\includegraphics[width=\linewidth]{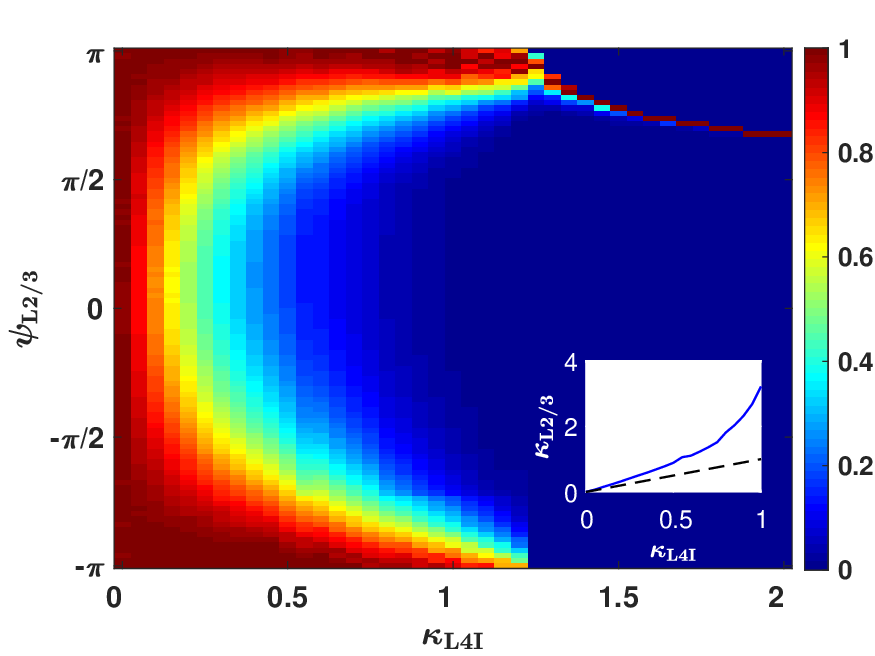} 
		\end{subfigure}\vspace{0.5cm}

		\caption{\textbf{STDP for $\kappa >0$.} (a) Synaptic weight dynamics. Each trace depicts the temporal evolution of a single synaptic weight, differentiated by color according to its preferred phase. (b)-(c) Dynamics of the order parameters: $\bar{w}$ (red) and $\tilde{w}$ (black) are shown in b, and $\psi$ is shown in c. (d)-(e) Polar histograms of the distributions of preferred phases of the L4I neurons (in d) with  $\kappa_{ \mathrm{L4I} }=0.6$ and  $\psi_{ \mathrm{L4I}}=0.25 \pi $ and the L2/3 neuron (in e) with  $\kappa_{ \mathrm{L2/3} }\approx1.2$ and  $\psi_{ \mathrm{L2/3}}\approx 2.3 \ \text{rad}$. (f) Histogram depicting the distribution of preferred phases of the downstream, L2/3, neuron over time. The characteristics of the distribution are identical to those in panel (e). The red dots show the inverse of the drift velocity of the L2/3 neuron, normalized such that the non-normalized data are multiplied by the ratio of the maximum value of the histogram to the maximum value of the non-normalized data. Each red dot is the average of 200 data points. (h) The distribution of the preferred phases of the downstream, L2/3, neurons, $\mathbf{\psi_{ \mathrm{L2/3} }}$, is shown in color code as a function of $\kappa_{\mathrm{L4I}}$. In each column, the un-normalized distribution: $\text{Pr}(\psi)/ \text{max} \{\text{Pr}(\psi)\}$, is 	depicted. The inset shows the distribution width downstream, in layer 2/3, $\kappa_{ \mathrm{L2/3} }$, as a function of the width of the distribution upstream, $\kappa_{\mathrm{L4I}}$ (blue). The identity line is shown (dashed black) for comparison. Here we used the following parameters: $N = 150$,  $\bar{\nu}=\nu/(2 \pi)= 10\text{hz}$, $D_{ \mathrm{L4I} } = 10 \text{hz}$,  $I_{ \mathrm{ex} } = 8 \text{hz}$, and $d=14 \text{ms}$. The temporally symmetric  STDP rule, \cref{eq:kernelSymmetric4}, was used with $\tau_-=20\text{ms}$, $\tau_+=50\text{ms}$,  $T_\pm=0\text{ms}$, $\mu=0.001$, and $\lambda=10^{-3}$. Initial conditions were random with a uniform distribution in the interval $[0.3, 0.7]$. All the data presented here were sampled from simulations with an integer number of cycles. For better visibility, the data presented in (a)-(c) shows a non-integer number of cycles.}
\end{figure*}

\section{\label{discussion}Discussion}
STDP can be viewed as a microscopic (unsupervised) learning rule. Typically, one pictures learning as a process that converges to an optimum. Thus, STDP dynamics is expected to relax to a fixed point or to a manifold attractor. However, empirical findings show that synaptic weights do not converge to a fixed point. Rather, synaptic weights remain volatile and vary by $\sim 50\% $ over a period of several days \cite{ziv2018synaptic,hazan2020activity}. Moreover, activity-independent plasticity on a similar order of magnitude has also been reported \cite{hazan2020activity}.

Thus, while activity-dependent plasticity (e.g.\ STDP) is considered to be an organizing force that pulls the system towards an attractor, activity-independent plasticity is thought to introduce noise into the learning dynamics. This noise will cause the synaptic weights to fluctuate around a point attractor or induce a random walk on a manifold attractor. Consequently, it has been argued that activity-independent plasticity can generate `drifting representations' \cite{driscoll2017dynamic,rule2020stable,kalle2021drifting,shimizu2021computational,wang2022tuning,driscoll2022representational,manz2023purely}. That is, the neural representation of an external stimulus will not be fixed in time.   

An example of a drifting representation was described by Driscoll and colleagues \cite{driscoll2017dynamic}. They showed that while the distribution of preferred stimuli across a population remained stable, the preferred stimuli of individual neurons varied over a period of several days. 

Here we showed that activity-dependent plasticity can also provide a mechanism that generates drifting representations. Moreover, this mechanism induces a distribution over time for the preferred stimuli of single neurons. This distribution is then translated into a distribution over the ensemble of neurons.

Our theory leads to several empirical predictions. The first and the most natural prediction is that preferred phases are dynamic and have a consistent drift velocity that is tightly related to the distribution of the preferred phases. In other words, drift velocity is expected to be minimal (maximal) at the most (least) common preferred phase.
 
Second, according to our theory, the distribution of preferred phases in the downstream population is governed by the STDP rule. A wide distribution, $\kappa_{ \mathrm{L2/3}} \approx 1$, implies that the fixed point solution of the STDP dynamics is unstable. Furthermore, the STDP rule also determines whether the mean phase of the downstream population will be advanced or delayed relative to the upstream. These observations yield quantitatively testable predictions on the temporal structure of the STDP rule of  L4I-to-L2/3 synapses.

Third, according to our theory, the distribution of the downstream layer is shifted and broadened due to STDP, when compared to the random pooling model, \hyperlink{fig 1}{Fig.\ 1(f)}. Consequently `turning off' STDP will cause the mean phase at the downstream population to shift towards that of the upstream, and the distribution is expected to become narrower. We do not yet know how to manipulate the STDP rule. However, STDP dynamics are driven by the product of the STDP rule and the cross-correlations of the neural activity. The cross-correlations can be modified by altering the sensory input to the system, e.g., by inducing artificial whisking. Such manipulations have been performed in the past, illustrating activity-dependent plasticity in the whisker system \cite{feldman2005map}. Thus, inducing artificial whisking at high frequencies is expected to weaken the STDP; see e.g.\ \hyperlink{Fig2}{Fig.\ 2(d)}, and consequently cause the distribution to shift its center and become narrower.  

Drifting representation has implications to our understanding of the neural code. Consider a linear readout that estimates the phase along the whisking cycle from the responses of L2/3 neurons. A linear estimator, for example, can be a weighted sum of the neural responses. An optimal linear estimator is a linear estimator with a choice of weights that maximizes the signal-to-noise ratio. Typically, for a linear readout the signal is embedded in the co-variation of the stimulus (whisking phase) and the neural responses \cite{salinas1994vector,shamir2014emerging}. However, if the neural representation of the stimulus is constantly changing, so is the `signal'. As a result, the weights of the optimal linear estimator will have to adapt constantly.

One alternative to consider is that the readout may not be optimal (at least not optimized to estimate the whisking phase)  \cite{shamir2014emerging}. As the distribution of preferred phases is not uniform, a linear readout that pools the neural responses with random weights will carry information on the whisking phase. Even though the accuracy of a random pooling readout is inferior to that of the optimal linear estimator, it is robust to drifting representation.
 
In our work, we made several simplifying assumptions. We ignored the possible contribution of recurrent connectivity within L2/3, which may 1) affect the pre-post cross-correlations, and 2) be plastic itself. In addition, the pre-post correlations were governed solely by rhythmic activity; however, a strong collective signal such as touch may also affect the STDP dynamics. Studying these effects was beyond the scope of the current work and will be addressed elsewhere. 

Nevertheless, learning in the central nervous system has been addressed empirically on two separate levels. On the microscopic level, the STDP rule has been investigated in preparations that lacked the functionality of the system. On the macroscopic level, functionality and behavior were studied in preparations that do not enable the investigation of the synaptic learning rules. The theory developed here bridges this gap, and draws direct links between these two levels that can thus serve as testable predictions for our hypothesis.

%

\section{\label{Supp}Supplementary materials}

\subsection{The cross-correlation}\label{correlderivation}

The cross-correlation between the pre- and post-synaptic firing is the driving force of STDP dynamics. In our model, the pre-post correlations are governed by the correlations within the upstream population. The cross-correlation between neurons $j$  and $k$ of the pre-synaptic population obey 

\begin{equation}\label{eq:correprepre4}
\begin{split}
\Gamma_{(j, k)}(\Delta t)& =  D^2  \Big(1  + 
\frac{\gamma^2}{2} \cos[ \nu \Delta t+\phi_{j}-\phi_{k}]\Big)+ 
 \\& \delta_{j k} D \delta(\Delta t).
\end{split}
\end{equation}

Using \cref{eq:postfiring3}, the cross-correlation between the $j$th L4I neuron and the downstream L2/3 neuron is given by 
\begin{equation}\label{eq:corrFull4}
\begin{split}
\Gamma_{j, \text{ post}}(\Delta t) &=I_{ex}D-\frac{1}{N}\sum_{j=1}^{N}w_j \Gamma_{i,j}(\Delta t-d)
\\&=
 I_{ex}D- \frac{D}{N}\delta(\Delta t-d) w_{j} 
-D^2 
\bigg( \bar{w}  + 
\\ 
&\frac{\gamma^2}{2} \tilde{w} \cos[ \nu (\Delta t-d) + \phi_{j}-  \psi ]\bigg),
\end{split}
\end{equation}
where the order parameters are defined in \cref{eq:w_bar,eq:w_tilde}.
In the limit of $N \gg 1$ the cross-correlation can be written as follows	
\begin{equation}\label{eq:corrFullN}
    \begin{split}
        \Gamma_{j, \text{ post}}(\Delta t) &=D D_\text{ post} (1+\frac{1}{2}\gamma \gamma_\text{ post} \cos[ \phi_{j}-  \phi_\text{ post}
        \\
    &+ \nu (\Delta t-d)])
    \end{split}
\end{equation}
where $ D_\text{ post},  \gamma_\text{ post}, \text{and}  \  \phi_\text{ post}$ are given in \cref{eq:PostStat}.

\subsection{STDP dynamics}\label{STDPderivation}
Substituting the pre-post cross-correlations in the large $N$ limit, \cref{eq:corrFullN}, into the STDP dynamics, \cref{eq:wdot4}, yields  
\begin{equation}\label{eq:wdotGeneral}
\begin{split}
    \frac{\dot{w}_i}{\lambda}&=D(I_{ex}-\bar{w}D)(f_+(w_i)-f_-(w_i))+\frac{D^2 \gamma^2}{2}\Big(\tilde{w}(\tilde{K}_+(\nu)\\
    & f_+(w_i) \cos[ \phi_{i}-\Omega_+  -  \phi_\text{ post}]-\tilde{K}_-(\nu)f_-(w_i)\\
    & \cos[ \phi_{i}-\Omega_-  -  \phi_\text{ post}]\Big),
\end{split}
\end{equation}
where $\tilde{K}_\pm(\nu)$, and $\Omega_\pm(\nu)$, are as defined in \cref{eq:Kfourier5}.

\subsection{Stability of the homogeneous fixed-points}\label{fpsstability}

In the homogeneous case, $\kappa = 0$, the STDP dynamics in the limit of large $N$, \cref{eq:wdotGeneral}, has two solutions, \cref{eq:steadystate4,eq:steadystate5}.

To examine the stability of the first solution, $w^*_1$, we  consider small perturbations around the fixed point $w=w^*_1+\delta w$, writing
\begin{equation}\label{eq:deltaw4Flac}
    \frac{\delta \dot{w}(\phi)}{\lambda}=\delta I(\phi)^+-\delta I(\phi)^-
\end{equation}
where
\begin{equation}\label{eq:deltaI}
\begin{split}
\\& \delta I(\phi)^\pm =\frac{\partial f_\pm(w_1^*)}{\partial w(\phi)} D(I_{ex}-D w_1^*)\delta w(\phi) 
\\
& -f_\pm(w_1^*)D^2  \big(\delta \bar{w}-\frac{ \gamma^2 \tilde{K}_\pm}{2} \cos(\phi-\psi-\nu d-\Omega_\pm)\delta \tilde{w} \big) .
\end{split}
\end{equation}
yielding
\begin{equation}\label{eq:deltawNEW}
	\begin{split}
&\frac{\delta \dot{w}(\phi)}{\lambda}=-g_0 \delta w(\phi)+ \Delta f D^2 \delta \bar{w}+\frac{D^2 \gamma^2}{2}\delta \tilde{w} ( \tilde{K}_- f_- \\ & \cos(\phi-\psi-\nu d-\Omega_-) - \tilde{K}_+ f_+ \cos(\phi-\psi-\nu d-\Omega_+)),
	\end{split}
\end{equation}	
with the notation $\Delta f= f_--f_+$ and 
\begin{equation}\label{eq:g0NggSupp}
    g_0 = \mu D^2 (\frac{I_{ex}}{w_1^*D}-1)\frac{w^{*\mu}_1}{1-w_1^*}=\mu  D^2 (\frac{I_{ex}}{D}-\frac{1}{2})2^{2-\mu}
\end{equation}		
Recalling that $\Delta f(w^*_1)=0$, the eigenvalue corresponding to uniform fluctuations is given by
\begin{equation}\label{eq:lambdabar1Supp}
    m_{u,1}=- g_0. 
\end{equation}	
Note that in this case the uniform eigenvalue does not depend on the learning rule.

The eigenvalue in the whisking direction is 
\begin{equation}\label{eq:lambdatilde1Supp}
	\begin{split}
    & m_{w,1}= m_{u,1}+\frac{1}{4} D^2 \gamma^2  (\tilde{K}_- f_-(w_1^*) \cos(\nu d+\Omega_-) -\\ & \tilde{K}_+ f_+(w_1^*) \cos(\nu d+\Omega_+))= m_{u,1}+\frac{1}{2^{2+\mu}} D^2 \gamma^2   \tilde{K} \cos( \alpha_0 ),
\end{split}
\end{equation}	
where $ \tilde{K}$ and $\alpha_0$ were defined in \cref{eq:alpha0}.


The corresponding eigenvalues around the balanced fixed point, \cref{eq:steadystate5}, obey
\begin{equation}\label{eq:lambdabar2Supp}
 m_{u,2}=-D^2\big( (1-w^*_2)^\mu -  (w^*_2)^\mu \big),
\end{equation}	
and  
\begin{equation}\label{eq:lambdatilde2Supp}
	\begin{split}
 m_{w,2}= &\frac{1}{4} D^2 \gamma^2  \big(\tilde{K}_- f_-(w_2^*) \cos(\nu d+\Omega_-) -\\ & \tilde{K}_+ f_+(w_2^*) \cos(\nu d+\Omega_+)\big).
\end{split}
\end{equation}	

\subsection{Derivation of drift velocity in the additive model}\label{phasedyn}
The phase of the synaptic weights profile, $\psi$, is given by \cref{eq:w_tilde} via the condition $  0 \leq \tilde{w} \in \mathbb{R}$, namely:
\begin{equation} \label{eq:Psi}
    0 = \int w(\phi) \sin ( \phi - \psi) \Pr(\phi) \text{d}\phi 
\end{equation}
where we take the continuum limit. Note that $ \text{Pr}(\phi)$ is the probability density of the preferred phases in the upstream (L4I) layer, \cref{eq:VM4}.

Taking the temporal derivative of \cref{eq:Psi} yields the dynamic equation for the phase, $\psi$,
\begin{equation}\label{eq:psidotcont}
    \dot{\psi}=\frac{1}{\tilde{w} } \int \dot{w}(\phi) \sin{(\phi-\psi)} \text{Pr}(\phi) \text{d}\phi.
\end{equation}

To utilize \cref{eq:psidotcont} we need a better understanding of the shape of the moving profile of synaptic weights, $w(\phi)$. Below, we first study the synaptic weights profile in the limit of small $\mu$, and derive a semi-phenomenological model for the moving profile. Next, we use a self-consistent argument to determine the parameters that characterize this profile. Then, we use \cref{eq:psidotcont} to compute the drift velocity.

\subsubsection{The semi-phenomenological model for small $\mu$ } 
\hyperlink{Fig3}{Fig.\ 3(h)} shows an example of a moving profile with a positive drift velocity and small $\mu$. The synaptic weights profile consists of four distinct regions. Two are saturated regions, in which the synapses are either fully depressed or fully potentiated. The other two are fronts; a preceding front and a receding front. 

In the saturated regions, $w(\phi) = 0$ or $w(\phi) = 1$, and due to the weight dependence of the STDP rule, $f_\pm (w)$, $\dot{w}(\phi) = 0$. As a result, the saturated regions do not contribute to the drift velocity; see \cref{eq:psidotcont}. In the interior of the fronts $w(\phi) \in (0,1)$, and in the limit of the additive model, $\mu \rightarrow 0$, $f_+(w) = f_-(w) = 1$. Consequently, in the fronts: 
\begin{equation}\label{eq:w_dot_front}
    \dot{w} (\phi) = \frac{1}{2} \lambda \tilde{w} D^2 \tilde{K} \gamma^2 \cos[ \alpha_0 + \psi -\phi ] , \text{ \ \ \ } \phi \text{ \ in fronts}.
\end{equation} 

The edges of the fronts, $\phi_{c_{1/2}}$, are given by the condition of the vanishing temporal derivative, $\dot{w} (\phi) =0$, in \cref{eq:w_dot_front}, yielding (compare to \cref{eq:muinvariant})
\begin{equation}\label{eq:phi_c12}
    \phi_{c_{1,2} } = \alpha_0 + \psi \pm \pi/2.
\end{equation}

\begin{figure}  \hypertarget{Fig5}{}
		\centering
		\begin{subfigure}[t]{1\textwidth} 
			\centering
			\includegraphics[width=\linewidth]{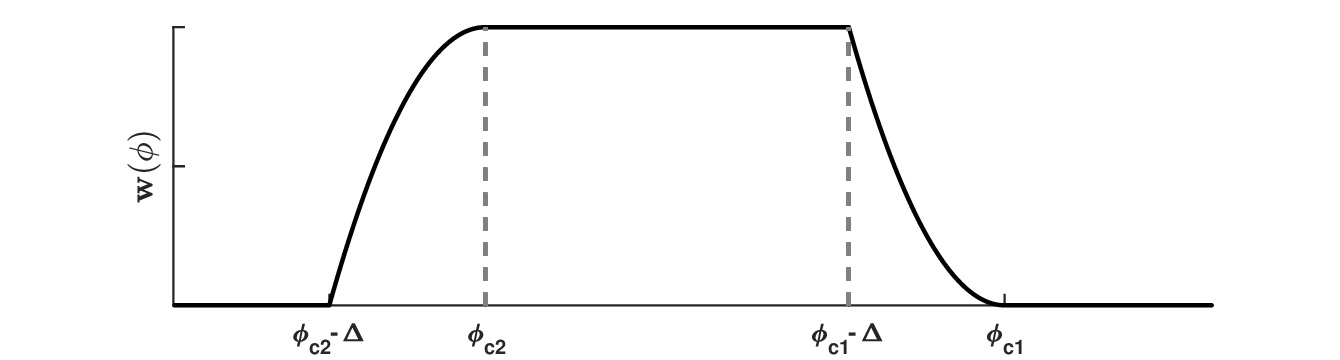} 
		\end{subfigure}
		\caption{\textbf{Synaptic weights profile.} The semi-phenomenological model for the moving profile of synaptic weights as a function of $\phi$, as described in \cref{eq:phenomenological}.}
\end{figure}

To compute the drift velocity we use the following semi-phenomenological model for the synaptic weights profile (see \hyperlink{Fig5}{Fig.\ (5)}):
\begin{equation} \label{eq:phenomenological}
    w(\phi) =  \left\{
    \begin{array}{ll}
        1 - \left( \frac{\phi - \phi_{c_1} }{ \Delta}\right)^2 & \phi \in (\phi_{c_2} - \Delta, \phi_{c_2}) \\
        1 & \phi \in (\phi_{c_2}, \phi_{c_1} - \Delta) \\
        \left( \frac{\phi - \phi_{c_1} }{ \Delta}\right)^2 & \phi \in (\phi_{c_1} - \Delta, \phi_{c_1}) \\
        0 & \text{otherwise} 
    \end{array}
    \right.
\end{equation}
Note that \cref{eq:phenomenological} must be taken with a grain of salt, as $\phi$ is an angular variable. \hyperlink{Fig3}{Figures 3(h)-3(i)} illustrates the agreement between our semi-phenomenological model and a moving profile obtained from simulating the STDP dynamics. The parameter $\Delta$ denotes the width of the front. 

The synaptic weights profile, in the limit of small $\kappa$, is fully defined by $\phi_{c_1}$, $\phi_{c_2}$ and $\Delta$. The profile determines $\psi$ via \cref{eq:w_tilde}, which in turn must be consistent with $\phi_{c_1}$ and $\phi_{c_2}$ as determined by \cref{eq:phi_c12}. This self-consistent constraint determines the width of the fronts, $\Delta$.

Below we determine the width of the front, $\Delta$, and use it to compute the drift velocity. First, in the homogeneous case, $\kappa = 0$. Then, we extend our results to a non-uniform distribution of preferred phases for $0 < \kappa \ll 1$, in a perturbative manner. 

\subsubsection{Uniform distribution} \label{phasedynISO}
In the case of a uniform distribution, $\kappa=0$, instead of using  \cref{eq:Psi} to determine $\psi$, we demand that $\psi$ is the `center of mass', $\psi_{\text{com}}$, of the synaptic weights profile:  
\begin{equation}\label{eq:psiCOM}
    \psi = \psi_{\text{com}} \equiv \frac{1}{\int w(\phi) \text{d}\phi} \int w(\phi) \phi \text{d}\phi.
\end{equation} 

Using \cref{eq:phenomenological}, the denominator of \cref{eq:psiCOM}
\begin{equation}\label{eq:wprofileAVG}
\begin{split}
    \int w(\phi)d\phi &=\int_0^{\Delta}  
    \left( \frac{\phi}{\Delta } \right)^2 d\phi+[\phi_{c_1}-\Delta -\phi_{c_2}]+ \\ & \int_0^{\Delta } \left(1- \left( \phi / \Delta \right)^2 \right) d\phi = \pi.
\end{split}
\end{equation}
The numerator of \cref{eq:psiCOM} gives 
\begin{equation}\label{eq:wprofilephi}
\begin{split}
    \int w(\phi)\phi d\phi &=\int_0^{\Delta}  \left( \frac{\phi}{\Delta}\right)^2 \phi d\phi+\int_{\phi_{c_2}}^{\phi_{c_1}-\Delta } \phi d\phi+ \\ & \int_0^{\Delta } \left(1-(\phi /\Delta)^2 \right)\phi d\phi=\pi(\alpha_0 + \psi -2\Delta/3).
\end{split}
\end{equation}
yielding from the self-consistency constrain, \cref{eq:psiCOM},
\begin{equation}\label{eq:deltaphi0}
    \Delta =\frac{3\alpha_0}{2}.
\end{equation}

Using the fact that the saturated regions do not contribute to the drift velocity,  \cref{eq:psidotcont} yields
\begin{equation}\label{eq:psidotCont1}
\begin{split}
    \frac{\dot{\psi}(t)}{\lambda} &= \frac{\gamma^2 D^2 \tilde{K}}{4\pi} \bigg\{ \int_{\phi_{c_2}-\Delta }^{\phi_{c_2}}\cos{(\alpha_0+\psi-\phi)}\sin{(\phi-\psi)} \text{d}\phi\\ &  \int_{\phi_{c_1}-\Delta }^{\phi_{c_1}}\cos{(\alpha_0+\psi-\phi)}\sin{(\phi-\psi)}\text{d}\phi \bigg\} = \\ & \frac{\gamma^2 D^2 \tilde{K}}{4} \big( 2 \Delta \sin[\alpha_0] + \cos[\alpha_0-2\Delta] - \cos[\alpha_0] \big),
    \end{split}
\end{equation}
substituting the value of $\Delta$, \cref{eq:deltaphi0}, yields  \cref{eq:Vdrift0}:
\begin{equation}\label{eq:psidotCont2}
    v_{ \mathrm{drift} }^{(0)} =\frac{\lambda \gamma^2 D^2 \tilde{K}}{4} \big(\cos(2\alpha_0) - \cos(\alpha_0)+3\alpha_0  \sin(\alpha_0)\big).
\end{equation}
Note that $v_{ \mathrm{drift} }^{(0)}$ does not depend on $\psi$. The superscript 0 denotes that this is the zero-order in $\kappa$. Below we now extend the calculation of the drift velocity in a perturbative manner in $ \kappa $. 

\subsubsection{Nonuniform distribution in the limit of small $\kappa$} \label{phasedynkappa}

To study the leading non-trivial order of the drift velocity, $v_{ \mathrm{drift} }$, in $\kappa$, 
we write
\begin{eqnarray}\label{eq:Delta_0_1}
   v_{ \mathrm{drift} }(\psi) = v_{ \mathrm{drift} }^{(0)} + \kappa v_{ \mathrm{drift} }^{(1)}(\psi) + \mathcal{O} (\kappa^2) ,
   \\
   \Delta(\psi) = \Delta^{(0)} + \kappa \Delta^{(1)}(\psi) + \mathcal{O} (\kappa^2),
\end{eqnarray}
where $v_{ \mathrm{drift} }^{(n)}$ and $\Delta^{(n)}$ ($n \in \mathbb{N} \cup \{ 0 \}$) are independent of $\kappa$. The drift velocity, $v_{ \mathrm{drift} }$, can be estimated from \cref{eq:psidotcont}, once the synaptic weights profile $w(\phi) $ is established, 
given the width of the front, $\Delta$.  

In the semi-phenomenological model, \cref{eq:phenomenological}, the synaptic weights profile is fully determined by $\psi$ (which determines $\phi_{c_{1,2}}$) and $\Delta$. The width of the front, $\Delta$, can be determined from \cref{eq:Psi}, yielding a first order in $\kappa$
\begin{equation}\label{eq:wIm}
    0 = L_0 + L_1,
\end{equation}
where
\begin{align}
    \label{eq:L0}
    & L_0= \int w(\phi, \Delta^{(0)} + \kappa \Delta^{(1)}) \sin(\phi-\psi) \text{d}\phi
    \\
    \label{eq:L1}
    &L_1= \kappa \int w(\phi, \Delta^{(0)} ) \sin(\phi-\psi) \cos(\phi) \text{d}\phi.
\end{align}
where for $0< \kappa \ll 1$, we approximated the probability density of preferred phases,  \cref{eq:VM4}, by 
\begin{equation}\label{eq:VMsmall}
    \Pr(\phi)\approx \frac{1+\kappa \cos(\phi)}{2 \pi}.
\end{equation}

In \cref{eq:L0,eq:L1} we explicitly denote the dependence of the synaptic weight profile on the width of the front. Note that as \cref{eq:L1} is $\mathcal{O} (\kappa)$, it is sufficient to take only the zero order in $\kappa$ for the width. 

The term, $L_0$ also contains zero order in $\kappa$, 
\begin{equation}\label{eq:L0kappa}
    L_0 =  L_0(\Delta^{(0)})+\kappa \Delta^{(1)}  \frac{\partial L_0}{\partial  \Delta} \biggr \rvert_{\Delta^{(0)}} + \mathcal{O} (\kappa^2 ). 
\end{equation}

To compute $ L_0(\Delta^{(0)})$ we consider the intervals in which $w(\phi)$ is nonzero:
\begin{equation}
     L_0=L_{0,0}+L_{0,1}+L_{0,2},
\end{equation}
where
\begin{align}
    \label{eq:L00}
    & L_{0,0}=\int_{\phi_{c_2}-\Delta }^{\phi_{c_1}-\Delta }\sin(\phi-\psi) \text{d}\phi,
    \\
    \label{eq:L01}
    &L_{0,1}=-\int_{\phi_{c_2}-\Delta }^{\phi_{c_2}} \frac{(\phi-\phi_{c_2})^2}{\Delta^2} \sin(\phi-\psi) \text{d}\phi,
    \\
    \label{eq:L02}
    &L_{0,2}=\int_{\phi_{c_1}-\Delta }^{\phi_{c_1}} \frac{(\phi-\phi_{c_1})^2}{\Delta^2} \sin(\phi-\psi) \text{d}\phi.
\end{align}
%
Thus, yielding
\begin{equation}\label{eq:L0sol}
     L_0(\Delta)=\frac{4}{ \Delta^2} \Big(\sin (\alpha_0 -\Delta )+ \Delta   \cos (\alpha_0 -\Delta )-\sin (\alpha_0 ) \Big).
\end{equation}
Note that the solution of \cref{eq:Psi} to order $ \kappa^0 $ is different than the `center of mass' calculation of \cref{eq:psiCOM}. Nevertheless, we find that $\Delta_{\mathrm{com}^{(0)}} = 3\alpha_0/2$ yields a good approximation of the solution of $L_0 (\Delta^{(0)}) =0$.

The first order in $\kappa$ of $L_0$ (see \cref{eq:L0kappa}) obeys
\begin{equation}\label{eq:L0kappaDer}
    \kappa \Delta^{(1)} \frac{\partial L_0}{\partial  \Delta }\biggr \rvert_{\Delta^{(0)}}=\kappa \Delta^{(1)} \frac{4}{ \Delta^{(0)}} \sin(\alpha_0-\Delta^{(0)}). 
\end{equation}
Substituting $\Delta^{(0)} \approx 3\alpha_0/2$ yields 
\begin{equation}\label{eq:L0kappafinal}
    \kappa \Delta^{(1)} \frac{\partial L_0}{\partial  \Delta }\biggr \rvert_{\Delta^{(0)} }= -\kappa \Delta^{(1)} \frac{8}{3 \alpha_0} \sin\big(\frac{\alpha_0}{2}\big). 
\end{equation}

The term $L_1$ is computed using the profile for $\kappa = 0$. Writing   \begin{equation}\label{eq:L1sol}
     L_1=L_{1,0}+L_{1,1}+L_{1,2},
\end{equation}
with
\begin{align}
    \label{eq:L10}
    & L_{1,0}= \kappa \int_{\phi_{c_2}-\Delta^{(0)}}^{\phi_{c_1}-\Delta^{(0)}} \sin(\phi-\psi) \cos(\phi) \text{d}\phi,
    \\
    \label{eq:L11}
    &L_{1,1}= -\kappa \int_{\phi_{c_2}-\Delta^{(0)} }^{\phi_{c_2}} \left( \frac{ \phi-\phi_{c_2} }{ \Delta^{(0)}} \right)^2 \sin(\phi-\psi)\cos(\phi) \text{d}\phi,
    \\
    \label{eq:L12}
    &L_{1,2}= \kappa \int_{\phi_{c_1}-\Delta^{(0)}}^{\phi_{c_1}} \left( \frac{ \phi-\phi_{c_1} }{\Delta^{(0)} } \right)^2 \sin(\phi-\psi) \cos(\phi) \text{d}\phi.
\end{align}

Substituting $\Delta^{(0)} \approx 3\alpha_0/2$ in \cref{eq:L10} one obtains
\begin{equation}\label{eq:L00sol}
    L_{1,0}=-\kappa\frac{ \pi \sin(\psi)}{2}.
\end{equation}
Noting that $ L_{1,1}+L_{1,2}=0$, \cref{eq:wIm} yields to first order in $\kappa$: 
\begin{equation}\label{eq:Delphi1}
    \Delta^{(1)} = -\frac{3\pi \alpha_0}{16 \sin\big(\frac{\alpha_0}{2}\big)}\sin(\psi).
\end{equation}
Hence, up to first order in $\kappa$, the width of the front is 
\begin{equation}\label{eq:Delphitaylor}
    \Delta(\psi) =\frac{3\alpha_0}{2} \left( 1 -\kappa \frac{ \pi  }{8 \sin\big(\frac{\alpha_0}{2}\big)}\sin(\psi) \right) .
\end{equation}

Expanding the probability density to first order in $\kappa$ in  \cref{eq:psidotcont}, and noting that in the limit of the additive rule, $\mu \rightarrow 0$, only the fronts contribute to the drift velocity, we obtain
\begin{equation}\label{eq:psidotkappa}
\begin{split}
     & v_{\mathrm{drift}}= \frac{ \lambda \gamma^2 D^2 \tilde{K}}{4\pi} \sum_{i=1,2} \int_{\phi_{c_i}-\Delta }^{\phi_{c_i}}(1+\kappa \cos(\phi))\cos{(\alpha_0+\psi-\phi) }\\ & \sin{(\phi-\psi)} \text{d}\phi 
    \end{split}
\end{equation}
where to first order in $\kappa$, $\Delta$ is given by \cref{eq:Delphitaylor}. \Cref{eq:psidotkappa} can be written as sum of three terms: 
\begin{equation}\label{eq:psidottotal}
     v_{\mathrm{drift}} = v_{\mathrm{drift}}^{(0)} + \kappa v_{\mathrm{drift}}^{(1a)} + \kappa v_{\mathrm{drift}}^{(1b)},
\end{equation}
where $ v_{\mathrm{drift}}^{(0)}$ is given by \cref{eq:psidotCont2} and 
\begin{equation}\label{eq:psi1a}
\begin{split}
    &v_{\mathrm{drift}}^{(1a)} =\frac{ \lambda \gamma^2 D^2 \tilde{K}}{4\pi} \sum_{i=1,2} \int_{\phi_{c_i}-\Delta \phi_0}^{\phi_{c_i}} \cos(\phi)\cos{(\alpha_0+\psi-\phi) }\\ & \sin{(\phi-\psi)} \text{d}\phi 
    \end{split}
\end{equation}
\begin{equation}\label{eq:psi1b}
\begin{split}
    & v_{\mathrm{drift}}^{(1b)} =\frac{\lambda}{\kappa}\frac{\gamma^2 D^2 \tilde{K}}{4\pi} \sum_{i=1,2} \int_{\phi_{c_i}-(\Delta^{(0)} + \kappa \Delta^{(1)} ) }^{\phi_{c_i} - \Delta^{(0)} } \cos{(\alpha_0+\psi-\phi) }\\ & \sin{(\phi-\psi)} \text{d}\phi 
    \end{split}
\end{equation}
Integrating the right hand side of \cref{eq:psi1a} over $\phi$ yields $ v_{\mathrm{drift}}^{(1a)} =0$, and in the limit of small $\kappa$, \cref{eq:psi1b}, yields
\begin{equation}\label{eq:psi1bsum}
\begin{split}
    & v_{\mathrm{drift}}^{(1b)} 
    \frac{ \lambda \gamma^2 D^2 \tilde{K}}{2\pi} \sin(\frac{3\alpha_0}{2}) \cos(\frac{\alpha_0}{2}) \Delta^{(1)}. 
    \end{split}
\end{equation}
Substituting \cref{eq:Delphi1} yields 
\begin{equation}\label{eq:psidottotalapprox}
\begin{split}
    & v_{\mathrm{drift}} (\psi) \approx v_{\mathrm{drift}}^{(0)} + \lambda \kappa \frac{\gamma^2 D^2 \tilde{K}}{2\pi} F(\alpha_0) \sin(\psi). 
\end{split}
\end{equation}
with 
\begin{align}\label{eq:Falpha0}
    F(\alpha_0) & =  - \frac{3 \pi \alpha_0 \sin( \frac{3 \alpha_0}{2} )}{
    16 \tan( \frac{ \alpha_0}{2} )
    },
\end{align}
as cited in \cref{eq:Vdrift1}.


\subsection{Numerical simulations \& pre-synaptic phase distributions} \label{Details of the numeric simulations4}

Scripts of the numerical simulations were written in Matlab. \hyperlink{Fig2}{Figure 2} was produced with Mathematica. Unless stated otherwise, the numerical results presented in this paper were obtained by solving \cref{eq:wdotGeneral} with the Euler method.
The cftool with the Nonlinear Least Squares method and a confidence level of $95 \%$ for all parameters was used for fitting the data presented in  \hyperlink{Fig4}{Figs.\ 4(d)-4(e)}.

Unless stated otherwise, we simulated the STDP dynamics in the mean field limit  without quenched disorder. To this end, the preferred phase, $\phi_k$, of the $k$th neuron in a population of $N$ pre-synaptic L4I neurons was set by the condition $ \int_{- \pi} ^{ \phi_k} \Pr ( \varphi) d  \varphi = k/N $.

\section*{\label{sec:ACKNOWLEDGMENTS}ACKNOWLEDGMENTS}
This research was supported by the Israel Science Foundation (ISF) grants number 300/16 and 624/22.

\bibliographystyle{unsrt}

\bibliography{MainPaper}
\end{document}